\documentclass[journal]{IEEEtran}
\usepackage{ifpdf}
\usepackage{cite}
\usepackage{algorithmic}
\usepackage[ruled,vlined]{algorithm2e}
\ifCLASSINFOpdf
   \usepackage[pdftex]{graphicx}

\else
 \usepackage[dvips]{graphicx}
\fi

\ifCLASSOPTIONcompsoc
    \usepackage[caption=false, font=normalsize, labelfont=sf, textfont=sf]{subfig}
\else
\usepackage[caption=false, font=footnotesize]{subfig}
\fi
\usepackage{titlesec}
\titlespacing*{\section}{0pt}{0.5\baselineskip}{0.5\baselineskip}
\titlespacing*{\subsection}{0pt}{0.5\baselineskip}{0.5\baselineskip}
\usepackage[cmex10]{amsmath}
\usepackage{epstopdf}
\usepackage{array}
\usepackage{flushend}
\usepackage{balance}
\usepackage{eqparbox}

\usepackage{fixltx2e}
\usepackage{color}
\usepackage{float}

\usepackage{stfloats}
\usepackage{url}
\usepackage{cite}
\usepackage{amsthm}

\newcommand*{\Scale}[2][4]{\scalebox{#1}{$#2$}}
\usepackage{amssymb}
\begin{document}

\title{Joint Beamforming Design for Multiuser MISO Downlink Aided by a Reconfigurable Intelligent Surface and a Relay 
\thanks{Part of this work is accepted to be presented at the IEEE International Conference on Communications (ICC workshop) 2021 \cite{obeed2021relay}.}}

\setlength{\columnsep}{0.21 in}

\author{Mohanad~Obeed~\thanks{ M. Obeed, and A. Chaaban are with the School of Engineering, University of British Columbia (UBC), Kelowna, British Columbia, Canada (email:  mohanad.obeed@ubc.ca, anas.chaaban@ubc.ca).}
 and Anas Chaaban,~\IEEEmembership{Senior Member,~IEEE,}}

\maketitle

\begin{abstract}
Reconfigurable intelligent surfaces (RIS) have drawn considerable attention recently due to their controllable scattering elements that are able to direct electromagnetic waves into desirable directions. Although RISs share some similarities with relays, the two have fundamental differences impacting their performance. To harness the benefits of both relaying and RISs, a multi-user communication system is proposed in this paper wherein a relay and an RIS cooperate to improve performance in terms of energy efficiency. Using singular value decomposition (SVD), semidefinite programming (SDP), and function approximations,  we propose different solutions for optimizing the beamforming matrices at the base-station (BS), the relay, and the phase shifts at the RIS to minimize the total transmit power subject to quality-of-service (QoS) constraints. The problem is solved in different cases when the relay operates in half-duplex and full-duplex modes and when the reflecting elements have continuous and discrete phase shifts. Simulation results are provided to compare the performance of the system with and without the RIS or the relay in both full-duplex and half-duplex modes, under different optimization solutions. Generally, the results show that the system with full-duplex relay and RIS cooperation outperforms all the other scenarios, and the contribution of full-duplex relay is higher than that of the RIS. However, an RIS performs better than a half-duplex relay when the required QoS is high. The results also show that increasing the number of RIS reflecting elements improves performance better in the presence of a relay than in its absence.   

\end{abstract}
\IEEEpeerreviewmaketitle

\section{Introduction}
The performance of wireless communication systems is affected by wave propagation phenomena such as  scattering, delays, reflections, and diffractions. Such phenomena makes the received signal consist of a superposition of multiple random, delayed, and attenuated copies of the transmitted signal, which impacts the achievable rate, energy efficiency, and coverage. Thus, adding some level of reconfigurability to the propagation medium can improve performance. Recently, a solution has been proposed to achieve this goal via deploying a reconfigurable intelligent surface (RIS), which is a two dimensional surface consisting of digitally-controllable reflecting elements \cite{alouiniR26,cuiR22, WuR21}. 

The elements of an RIS can be controlled to achieve a specified objective such as strengthening the received signal or weakening the received interference. Specifically, the RIS elements can control the phase shifts between the incident and reflected waves so that the received waves add up constructively or destructively at a receiver. In addition to the ability of RISs to configure the radio environment, they have other desired properties 
 such as: i) easy installation,  ii) low cost, iii) passive elements, and iv) low energy consumption. In addition, RIS can perform its function without the need for sophisticated radio-frequency circuits and digital signal processing.  All these features allow an RIS to realize a smart radio environment that can be programmed to improve customized objectives such as spectral efficiency, energy efficiency, secrecy capacity, and coverage probability. This gives RISs the potential to play a role in sixth generation (6G) wireless networks.

In the literature, several papers investigate how the RIS phase shifts can be optimized along with the beamforming vectors at the BS, with the goal of improving communication performance. For instance, an RIS can significantly improve the sum-rate in a multiple-input-multiple-output (MIMO) single user or multiuser system \cite{9110889}, increase the energy efficiency of the communication system \cite{WuR1, huang2019reconfigurable}, decrease inter-cell interference \cite{9090356}, enhance the secrecy rate \cite{8723525}, maximize the minimum signal-to-interference and noise ratio (SINR) \cite{9087848}, and extend the coverage of millimetre wave communications \cite{9226616}. Interestingly, as shown numerically in \cite{WuR14}, the signal-to-noise ratio (SNR) of the RIS-assisted system with $N=100$ reflecting elements is improved by 10 dB compared to the non-RIS-assisted systems, and the SNR increases proportional to $N^2$ as $N$ increases. The authors in \cite{9194749, 9384319} show that supporting wireless networks with RIS significantly contribute in improving the sum-rate even if the phase shifts of the reflecting elements are randomly selected.

Although RISs can be used to improve the end-to-end channel quality, their main limitation is that the magnitude of the reflected channel (base-station (BS) to RIS to user) is weak compared to the direct channel between the BS and the user, since this channel is the product of the BS-RIS channel and the RIS-user channel, and this phenomena is called the double fading effect. This limitation can be overcome if the RIS is replaced by a relay node (both of which share the forwarding functionality), at the expense of additional energy consumption and complexity at the relay. The authors in \cite{HuangR3} compare RIS and relay-assisted systems and show that the RIS-assisted system is able to provide up to 300\% higher energy efficiency compared to the use of a multi-antenna angular amplify-and-forward (AF) relaying system. This large gain is due to the fact that an RIS consumes very little power to forward the impinging signal contrary to a relay. However, the authors in \cite{sonR25} show that an RIS-assisted system cannot provide an SNR higher than that in a massive MIMO decode-and-forward (DF) relay-assisted system for any value of $N$, again at the cost of more power consumption. This is due to fact that the end-to-end channel of the DF relay is much better than the reflected channel through the RIS.  
Authors of \cite{RIS_Relay} discuss similarities and differences between RISs and relays, and argue that an RIS-assisted system outperforms a relay-assisted one in terms of data rates when the RIS is sufficiently large. 

From the previous works, we can conclude that the RIS-assisted system  performs better than the AF relay-assisted system, but worse than a DF relay-assisted one when the relay is equipped with a massive number of antennas. 
In general, each of the DF relay and the RIS have their own pros and cons in terms of energy efficiency, hardware and software complexity, range, etc., and it is interesting to combine the two in a system to harness both gains from the RIS and the DF relay. 
We note that while all the aforementioned papers consider a system with an RIS or a relay, but not both, \cite{ying2020relay,9225707} study a system that consists of both an RIS and a relay. However, \cite{ying2020relay,9225707} focus on a single-user system with a single-antenna transmitter and receiver and maximize the achievable rate. Since a multi-antenna multi-user system is more practical (and challenging),  in this paper, we study a multi-user system employing a multi-antenna BS, a multi-antenna  DF relay, and an RIS. Our goal is to optimize the parameters of the system (beamforming matrices and reflection phases) to improve the multi-user system energy efficiency. 

\subsection{Contribution}
The main contributions of the paper can be summarized as follows:
\begin{itemize}
\item We describe transmissions schemes for a system consisting of a multi-antenna BS, a multi-antenna DF relay, an RIS, and multiple users, and we express the achievable rate of the schemes under both full-duplex and half-duplex relay operation.
\item We formulate optimization problems to minimize the total power under QoS constraints by jointly designing the beamforming matrices at the BS and the relay and the phase shifts at the RIS. We propose alternating optimization methods that solve for the beamforming parameters and the phase shifts iteratively. The problem is solved in different cases when the relay operates in half-duplex and full-duplex modes and when the reflecting elements have continuous and discrete phase shifts.
\item We compare the performance of the proposed system with system assisted by a relay-only and an RIS-only  as benchmarks. We also study the impact of increasing the number of reflecting elements on performance, when a relay is present or absent. 
\end{itemize}
\subsection{Discussion}
To design the beamforming matrices, we first notice that the optimization of the BS and relay beamforming matrices is coupled in the formulated optimization problem. Thus, we first decouple these optimizations relying on the fact that the equivalent BS-user channels are much weaker than the equivalent relay-user channels. We then propose a solution using the singular value decomposition (SVD) and uplink-downlink duality approaches. 
For both full-duplex and half-duplex scenarios,  we propose to place the RIS in the vicinity of the relay to enable better utilization of the reflecting elements in both hops (BS-to-relay and relay-to-users). This leads to optimize $2L$ phase shifts in the half-duplex mode and $L$ phase shifts in the full-duplex mode. 
For optimizing the phase-shifts, we propose two solutions for each scenario (full and half-duplex) when the phase-shifts are continuous and two other solutions for the discrete phase shifts case. The solutions are based on SDP, signal energy maximization with fixed-point iteration, quantization, and successive refinement approach. Some of these approaches are proposed to achieve high performance and others to achieve low complexity.

 We demonstrate numerically that the relay-RIS cooperation significantly reduces the required transmit power compared to the other systems. The results show that the contribution of the full-duplex relay in enhancing the performance is higher than that of the RIS, whose contribution in turn is better than the half-duplex relay when the required data rates (QoS) at the users is high. We also demonstrate numerically that increasing the number of reflecting elements has higher impact in the presence of a relay more than in its absence. In particular, increasing the number of reflecting elements from 20 to 180 reduces the required power by 0.28 dBm if there is no relay (RIS-only) and by 3.94 dBm if there is a full-duplex relay with 5 antennas (the proposed system). This means that the presence of a relay increases the effectiveness of RIS.


Next, we present the system model, describe the transmission scheme, and characterize the achievable rates. We optimize the half-duplex system in Sec. \ref{Sec:Form} and the full-duplex one in Section \ref{Sec:Full}. Special cases and baseline approaches are provided  in Section \ref{Sec:Spec}. We provide numerical evaluations of the performance of the schemes in Sec. \ref{Sec:Sim}. Finally, we conclude in Sec. \ref{Sec:Conc}.

\section{System Model}

\begin{figure}[!t]
\centering
\includegraphics[width=3.4in]{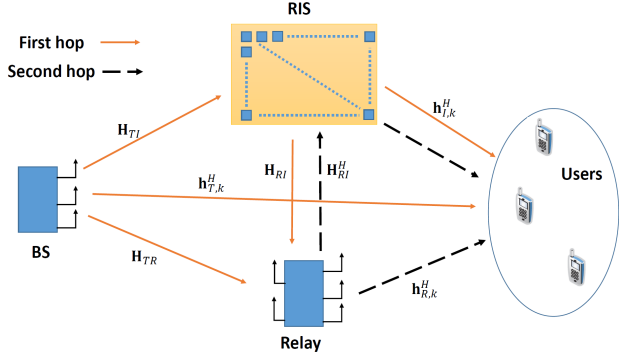}
\caption{System model consisting of a base-station, a relay, RIS, and $K$ users, assuming half-duplex operation. The solid arrows indicate signal flow in the first phase, while dashed arrows indicate the signal flow in the second phase.}
\label{SM}
\end{figure}

We consider a downlink transmission system consisting of $K$ single-antenna users, a DF relay with $N$ antennas, a BS with $M$ antennas, and an RIS with $L$ elements (Fig. \ref{SM}). The RIS is located near the relay to make it useful in both hops for reflecting signals from the BS to the relay and from the relay to the user. Moreover, the RIS and relay are assumed to be closer to the users than to the BS.  We consider two cases where the relay either operates in a half-duplex or a full-duplex mode. In addition, we consider two cases for the RIS where the phase shifts are continuous or discrete.

The BS wants to transmit a message with rate $R_k$ (bits per symbol) to user $k$ for all $k=1,\ldots,K$. It encodes the messages into codewords $s_k(1),\ldots,s_k(T)$ of length $T$ symbols. The symbols $s_k(t)$ follow a circularly symmetric complex Gaussian distribution with zero mean and unit variance. The BS then combines symbols from all codewords into a transmit signal $\mathbf{x}(t)$ which is sent to the relay and the users as described next for the half-duplex and full-duplex modes, in the next time slot if the relay is operated in a half-duplex mode, or in the same time slot if it is operated in a full-duplex mode. Next, we present the signal model and the achievable rates in half-duplex and full-duplex modes. 



\subsection{Signal Model: Half-Duplex Mode}\label{sec:HD}

Here, we present the signal model and the achievable rates when the system works under the half-duplex constraint. Under such assumption, signals are transmitted in two phases. In phase 1, the BS sends messages intended to users $1,\ldots,K$ encoded with rates $R_1,\ldots,R_K$ to the relay, and the relay decodes these messages. In phase 2, the relay forwards the messages to the users, and the users combine the received signals from both phases and decode their desired messages. Note that due to this, although the encoding rate is $R_k$ (bit/symbol), the overall rate achieved by user $k$ is $R_k/2$ (bits per transmission) since each symbol is transmitted over two transmissions (phase 1 and phase 2). The role of the RIS is to reflect the transmitted signal from the transmitter to the relay and the users in the first phase, and from the relay to the users in the second phase. 

The two phases of the transmission process, each of length $T$ symbols where $T$ is the codeword length, are explained next.

\subsubsection{Phase 1}
In the first phase, the BS transmits a precoded signal $\mathbf{x}(t)$ that can be expressed as follows
\begin{equation}
\mathbf{x}(t)=\sum_{k=1}^K \mathbf{w}_ks_k(t), \ t=1,\ldots,T,
\end{equation}
where $\mathbf{w}_k \in \mathbb{C}^{M\times 1}$ is the bemaforming vector assigned to user $k$ and $s_k(t)$ is a symbol of the length-$T$ codeword to be sent to user $k$. The symbols $s_k(t)$ have unit power, and the transmit power of the BS in this phase is $\frac{1}{2}\sum_{k=1}^K \|\mathbf{w}_k\|^2$, where the $\frac{1}{2}$ is due to the half-duplex operation since the BS is active half of the time.

Each user receives a superposition of the transmitted signal through the direct links and the reflected links via RIS, given by 
\begin{equation}
y_{k,1}(t)=(\mathbf{h}_{I,k}^H\mathbf{\Theta}_1\mathbf{H}_{TI}+\mathbf{h}_{T,k}^H)\sum_{i=1}^K\mathbf{w}_is_i(t)+n_k(t),
\end{equation}
$t=1,\ldots,T$, where $\mathbf{H}_{TI}\in\mathbb{C}^{L\times M}$ and $\mathbf{h}_{T,k}^H\in\mathbb{C}^{1\times M}$ are the channel coefficient from the BS to the RIS, and the BS to the user $k$, respectively, $\mathbf{h}_{I,k}^H\in\mathbb{C}^{1\times L}$ is the channel vector from the RIS to the user $k$,  $\mathbf{\Theta}_1=\text{diag}[e^{j\theta_{1,1}}, e^{j\theta_{1,2}},\ldots,e^{j\theta_{1,L}}]$, $\theta_{1,\ell}$ is the phase shift of the $\ell$th RIS element in the first phase, and $n_k(t)$ is additive white Gaussian noise (AWGN) with zero mean and variance $\sigma^2$, independent through $t$. The user will combine this signal with the received signal in the second phase before decoding.

The received signal at the relay in this phase is given by
\begin{equation}
\mathbf{y}_{r}(t)= (\mathbf{H}_{TR}+\mathbf{H}_{IR}\mathbf{\Theta}_1\mathbf{H}_{TI})\sum_{i=1}^K\mathbf{w}_is_i(t)+\mathbf{n}_R(t),
\end{equation}
$t=1,\ldots,T$, where $\mathbf{H}_{TR}\in\mathbb{C}^{N\times M}$, $\mathbf{H}_{IR}\in\mathbb{C}^{N\times L}$ are the channel coefficient matrices from the BS to the relay and from the RIS to the relay, respectively, and $\mathbf{n}_R(t)$ is the AWGN vector at the relay with zero mean and covariance matrix $\sigma^2\mathbf{I}$, independent through $t$.

The relay uses a decode-and-forward scheme. It decodes the users' codewords from $\mathbf{y}_{r}(t)$, $t=1,\ldots,T$, which imposes a rate constraint $\sum_{k=1}^K R_k \leq R_{R}$ where
\begin{equation}
\label{Rr}
R_{R}=\log_2\vert \mathbf{I}+\frac{1}{\sigma^2}\mathbf{H}_{TIR}\mathbf{W}\mathbf{W}^H
\mathbf{H}_{TIR}^H \vert,
\end{equation}
where $\mathbf{W}=[\mathbf{w_1}\ \mathbf{w_2},\ldots, \mathbf{w_k}]$ and $\mathbf{H}_{TIR}=\mathbf{H}_{TR}+\mathbf{H}_{IR}\mathbf{\Theta}_1\mathbf{H}_{TI}$. After decoding, the relay knows $s_k(t)$ for all $k=1,\ldots,K$ and $t=1,\ldots,T$, which can be assumed error free if $T$ is large enough and $\sum_{k=1}^K R_i\leq R_R$ is satisfied.

\subsubsection{Phase 2}
After decoding, the relay multiplies the decoded codewords by vectors $\mathbf{u}_k$ to beamform them to the RIS and the users. The relay sends $\sum_{k=1}^K\mathbf{u}_ks_k(t)$ leading to a transmit power $\frac{1}{2}\sum_{k=1}^K\|\mathbf{u}_k\|^2$.

The phase shifts at the RIS can be chosen to be different from the ones used in first phase. Therefore, the received signal at the user $k$ in this phase becomes 
\begin{equation}
y_{k,2}(t)=(\mathbf{h}_{I,k}^H\mathbf{\Theta}_2\mathbf{H}_{IR}^H+\mathbf{h}_{R,k}^H)\sum_{i=1}^K\mathbf{u}_is_i(t)+n_2(t),
\end{equation}
$t=1,\ldots,T$, where $\mathbf{h}_{R,k}^H\in\mathbb{C}^{1\times N}$ is the channel vector from the relay to the user $k$, $\mathbf{\Theta}_2=\text{diag}[e^{j\theta_{2,1}}, e^{j\theta_{2,2}},\ldots,e^{j\theta_{2,L}}]$ is the phase shift matrix in phase 2 with $\theta_{2,\ell}$ being the phase shift applied by the $\ell$th RIS element, and $n_2(t)$ is AWGN with zero mean and variance $\sigma^2$, independent through $t$. Note that we assume channel reciprocity between the relay and the RIS, and hence the relay-RIS channel is the Hermitian of the RIS-relay channel. 

User $k$ uses maximum ratio combining (MRC) to combine $y_{k,1}(t)$ and $y_{k,2}(t)$ before decoding $s_k(t)$, $t=1,\ldots,T$. Decoding is reliable if $R_k\leq\bar{R}_k$ where 
\begin{equation}
\bar{R}_{k}= \log_2(1+\gamma_{k,1}+\gamma_{k,2}),
\end{equation}
where $\gamma_{k,1}$ and $\gamma_{k,2}$ are the received signal-to-noise ratios (SNRs) at user $k$ from the first and second phases, respectively, given by
\begin{align} 
\gamma_{k,1}&=\frac{\vert(\mathbf{h}_{I,k}^H\mathbf{\Theta}_1\mathbf{H}_{TI}+\mathbf{h}_{T,k}^H) \mathbf{w}_k\vert^2}{\sum_{i\neq k} \vert (\mathbf{h}_{I,k}^H\mathbf{\Theta}_1\mathbf{H}_{TI}+\mathbf{h}_{T,k}^H)\mathbf{w}_i\vert^2+\sigma^2}\\
\gamma_{k,2}&=\frac{\vert(\mathbf{h}_{I,k}^H\mathbf{\Theta}_2\mathbf{H}_{IR}^H+\mathbf{h}_{R,k}^H) \mathbf{u}_k\vert^2}{\sum_{i\neq k} \vert (\mathbf{h}_{I,k}^H\mathbf{\Theta}_2\mathbf{H}_{IR}^H+\mathbf{h}_{R,k}^H)\mathbf{u}_i\vert^2+\sigma^2}.
\end{align}

The goal is to minimize the transmit power while ensuring QoS constraints at the users. This problem is discussed in section \ref{Sec:Form}. Next, we describe the scheme and the achievable rate in the full-duplex case. 

\subsection{Signal Model: Full-Duplex Mode}\label{sec:FD}
Here, we model the signal and present the achievable rates when the relay operates in full-duplex mode. 
We divide the transmission to $B$ blocks, each of length $T$ symbols. At time instant $t$ of block $m$, 
the transmitted signal from the BS is given by
\begin{equation}
\mathbf{x}_m(j)=\sum_{k=1}^K \mathbf{w}_ks_{k}^m(t),
\end{equation} 
where $s_k^m(t)$ is the $t$th symbol of the codeword of user $k$ corresponding to block $m$. In the same block, the relay transmits the $t$th signal of the $m-1$ block and it is given by 
\begin{equation}
\mathbf{x}_{m}(j)=\sum_{k=1}^K \mathbf{u}_ks_{k}^{m-1}(t).
\end{equation}
Hence, the received signal at the relay at $t$ is given by
\begin{multline}
\mathbf{y}_{R,m}(t)=(\mathbf{H}_{IR}\mathbf{\Theta}\mathbf{H}_{TI}+\mathbf{H}_{TR})\sum_{k=1}^K \mathbf{w}_ks_{k}^{m}(t)\\
+(\mathbf{H}_{IR}\mathbf{\Theta}\mathbf{H}_{IR}^H+\mathbf{H}_{RR})\sum_{k=1}^K \mathbf{u}_ks_{k}^{m-1}(t)+\mathbf{n}(t),
\end{multline}  
where $\mathbf{H}_{RR}$ is the self-interference channel matrix from the relay to the relay. Aligned with our assumption, which assumes that each node is able to estimate the CSI perfectly, the relay is able to cancel this self-interference. Note that this self-interference cancellation can be achieved in practice as described in \cite{7105651}.

Therefore, the achievable rate at the relay is given by 
\begin{equation}
R_{R}^{FD}=\log_2|\mathbf{I}+\frac{1}{\sigma^2}\mathbf{H}_{TIR}\mathbf{W}\mathbf{W}^H\mathbf{H}_{TIR}^H|,
\end{equation}
where $\sum_k R_k\leq R_R^{FD}$ must be satisfied.
The received signal at the user $k$ is given by 
\begin{multline}
y_{k}^m(t)= (\mathbf{h}_{I,k}^H\mathbf{\Theta}\mathbf{H}_{IR}^H+\mathbf{h}_{R,k})\sum_{k=1}^K \mathbf{u}_ks_{k}^{m-1}(t)\\
+(\mathbf{h}_{I,k}^H\mathbf{\Theta}\mathbf{H}_{TI}+\mathbf{h}_{T,k})\sum_{k=1}^K \mathbf{w}_ks_{k}^m(t)+n(t).
\end{multline}
Since we assume that the channel from the relay to user $k$ $\forall k$ is better than the channel from the BS to the same user, it would be better for each user to decode $s_{k,j}^{m-1}$ while treating   $s_{k,j}^{m}$ as noise. Therefore, the achievable rate at the user $k$ is given by 
\begin{equation}
\label{Rkfd}
R_{k}^{FD}= \log_2\bigg(1+\frac{\vert\mathbf{h}_{IR,k}^H \mathbf{u}_k\vert^2}{\sum_{i\neq k}^K \vert \mathbf{h}_{IR,k}^H\mathbf{u}_i\vert^2+ \sum_{i=1}^K\vert \mathbf{h}_{TI,k}^H\mathbf{w}_i\vert^2+\sigma^2}\bigg), 
\end{equation} 
where $\mathbf{h}_{IR,k}^H=\mathbf{h}_{I,k}^H\mathbf{\Theta}\mathbf{H}_{RI}+\mathbf{h}_{R,k}^H$ and $\mathbf{h}_{TI,k}^H=\mathbf{h}_{I,k}\mathbf{\Theta}\mathbf{H}_{TI}+\mathbf{h}_{T,k}$, i.e., we must have $R_k\leq R_k^{FD}$.
Our goal is to minimize the transmit power while ensuring that the required QoS at the users are achieved. This problem will be formulated in Sec \ref{Sec:Full}.

\subsection{Channel Model}
 We assume that the BS-RIS and BS-relay channels have a line-of-sight (LoS) component which can be ensured by proper placement of the relay and RIS, while the channels to the users are non LoS. The channels maintain their value during a coherence interval and change independently between intervals (block fading). We also assume that the channel state information (CSI) of all links is known at the BS and the relay. Next, we formulate the optimization problem for minimizing the total power subject to QoS constraints when the relay operates in a half-duplex mode.  

\section{Problem Formulation And Proposed Algorithms: Half-Duplex Mode} \label{Sec:Form}
Our goal is to investigate the gains achieved by combining an RIS and a relay in terms of energy efficiency. Through this, we aim to study the impact of optimizing the beamforming matrices at the BS and the relay and the phase shifts at the RIS on the total transmit power under the QoS constraints, and to draw insight which can be useful for network design. To this end, we formulate the optimization problem as minimizing the power at the BS and the relay under QoS constraints, then we propose different solutions for the formulated problem. 

The QoS constraint is given by a threshold rate $R_{th}$ so that each user can achieve this rate. This implies that the achievable rate under half-duplex operation $R_k/2$ has to be larger than $R_{th}$. The problem can thus be formulated as follows
\begin{subequations}
\label{OP1}
\begin{eqnarray}
\label{OP1a}
&\displaystyle\min_{\mathbf{W}, \mathbf{\Theta}_1, \mathbf{\Theta}_2, \mathbf{U}}& \frac{1}{2}\sum_{k=1}^K\|\mathbf{w}_k\|^2+\frac{1}{2}\sum_{k=1}^K\|\mathbf{u}_k\|^2 \\
\label{OP1b}
&\text{s.t.}&   R_{R}\geq 2KR_{th},\\
\label{OP1c}
&& \bar{R}_{k}\geq 2R_{th},\ \ k=1,\ldots, K,\\
\label{OP1d}
&&\mathbf{\Theta}_1, \mathbf{\Theta}_2 \in \mathcal{F},
\end{eqnarray}
\end{subequations} 
where $\mathbf{U}=[\mathbf{u}_1,\ldots,\mathbf{u}_K]$, and $\mathcal{F}$ is the set of the feasible phase shifts of the reflecting elements. The objective function \eqref{OP1} is the total transmit power at the BS and the relay. Constraints \eqref{OP1b} and \eqref{OP1c} guarantees that the required QoS at users is achievable in both the first and second phases, respectively.

Note that problem \eqref{OP1} is nonconvex and the coupling between the variables $\mathbf{W}$, $\mathbf{U}$, $\mathbf{\Theta}_1$, and $\mathbf{\Theta}_2$ makes the problem more difficult. Moreover, even if the variables $\mathbf{\Theta}_1$ and $\mathbf{\Theta}_2$ are given, constraint \eqref{OP1c} is still nonconvex. Hence, there is no standard solution that finds the global optimum of this optimization problem. However, in what follows, we propose two efficient solutions based on alternating optimization, singular value decomposition (SVD), and uplink-downlink duality. We first propose a solution for both $\mathbf{W}$ and $\mathbf{U}$ when $\mathbf{\Theta}_1$ and $\mathbf{\Theta}_2$ are given. Then we propose solutions for $\mathbf{\Theta}_1$ and~$\mathbf{\Theta}_2$ under the assumption that the phases are continuous. After that, we propose two solutions for the discrete phase shift case.


\subsection{Optimizing $\mathbf{W}$ and $\mathbf{U}$}
For given $\mathbf{\Theta}_1$ and $\mathbf{\Theta}_2$, problem \eqref{OP1} reduces to 
\begin{subequations}
\label{OP11}
\begin{eqnarray}
&\displaystyle\min_{\mathbf{W}, \mathbf{U}}& \frac{1}{2}\sum_{k=1}^K\|\mathbf{w}_k\|^2+\frac{1}{2}\sum_{k=1}^K\|\mathbf{u}_k\|^2 \\
\label{OP11b}
&\text{s.t.}&   R_{R}\geq 2KR_{th},\\
\label{OP11c}
&& \bar{R}_{k}\geq 2R_{th},\ \ k=1,\ldots, K,
\end{eqnarray}
\end{subequations} 
Constraints \eqref{OP11c} still makes the problem nonconvex. Hence, we propose a suboptimal, yet efficient, solutions as follows.

\subsubsection{SVD and Uplink-Downlink Duality} 
Due to the assumption that the relay and RIS are much closer to the users than the BS, the BS-user channel gains in the first phase are weak compared to relay-user channel gain in the second phase. This means that $\gamma_{k,1}\ll\gamma_{k,2}$ and we can neglect $\gamma_{k,1}$ at the expression $R_K$. Since $\gamma_{k,2}$ is independent of W, we can ignore \eqref{OP11c} when we optimize $\mathbf{W}$. In other words, since optimizing $\mathbf{W}$ would produce a higher impact on $R_R$ than on $\bar{R}_k \ \forall k$, we design $\mathbf{W}$ to achieve constraint \eqref{OP11b} while ignoring \eqref{OP11c}.   
With that said, problem \eqref{OP11} can be separated into two optimization problems, one for $\mathbf{W}$ given by
\begin{subequations}
\label{OP2}
\begin{eqnarray}
&\displaystyle\min_{\mathbf{W}}& \frac{1}{2}\sum_{k=1}^K\|\mathbf{w}_k\|^2 \\
\label{OP2b}
&\text{s.t.}&   R_{R}\geq 2KR_{th},
\end{eqnarray}
\end{subequations}  
and the other for $\mathbf{U}$ with a given $\mathbf{W}$ given by
\begin{subequations}
\label{OP45}
\begin{eqnarray}
&\displaystyle\min_{\mathbf{U}}& \frac{1}{2}\sum_{k=1}^K\|\mathbf{u}_k\|^2 \\
\label{OP45c}
&\text{s.t.}& \frac{\vert(\mathbf{h}_{I,k}^H\mathbf{\Theta}_2\mathbf{H}_{RI}+\mathbf{h}_{R,k}^H) \mathbf{u}_k\vert^2}{\sum_{i\neq k} \vert (\mathbf{h}_{I,k}^H\mathbf{\Theta}_2\mathbf{H}_{RI}+\mathbf{h}_{R,k}^H)\mathbf{u}_i\vert^2+\sigma^2}\geq \eta_k,
\nonumber\\
&&\ \ \ \ \ \ \  k=1,\ldots, K,
\end{eqnarray}
\end{subequations} 
where   $\eta_k=2^{2R_{th}}-1-\frac{\vert(\mathbf{h}_{I,k}^H\mathbf{\Theta}_1\mathbf{H}_{TI}+\mathbf{h}_{T,k}^H) \mathbf{w}_k\vert^2}{\sum_{i\neq k} \vert (\mathbf{h}_{I,k}^H\mathbf{\Theta}_1\mathbf{H}_{TI}+\mathbf{h}_{T,k}^H)\mathbf{w}_i\vert^2+\sigma^2}$.
We start by focusing on \eqref{OP2}. The solution of \eqref{OP2} can be achieved using SVD with water filling. Specifically, by defining $\mathbf{H}_{TIR}=\mathbf{H}_{TR}+\mathbf{H}_{IR}\mathbf{\Theta}_1\mathbf{H}_{TI}=\bar{\mathbf{V}}\mathbf{\Lambda}^{\frac{1}{2}}\tilde{\mathbf{V}}^H$, where $\bar{\mathbf{V}}$ and $\tilde{\mathbf{V}}$ are the matrices of singular vectors and $\mathbf{\Lambda}$ is the matrix of singular values, the solution of $\mathbf{W}$ is given by 
\begin{equation}
\label{SVD}
\mathbf{W}=\tilde{\mathbf{V}}\bar{\mathbf{\Xi}}^{\frac{1}{2}},
\end{equation}
where $\bar{\mathbf{\Xi}}=\text{diag}[P_1, P_1,\ldots,P_K]$ is the power allocation matrix. Using \eqref{SVD}, problem \eqref{OP2} reduces to 
\begin{subequations}
\label{OP3}
\begin{eqnarray}
&\displaystyle\min_{\mathbf{P}}& \frac{1}{2}\sum_{k=1}^K P_k \\
\label{OP3b}
&\text{s.t.}&   \sum_{k=1}^K\log_2(1+\frac{1}{\sigma^2}P_k\lambda_k)\geq 2KR_{th},
\end{eqnarray}
\end{subequations}     
where $\lambda_k$ is the $k$th eigenvalue of the matrix $\mathbf{H}_{TIR}\mathbf{H}_{TIR}^H$. Problem \eqref{OP3} is convex and can be solved using the water filling. The optimal $P_k$ is given by 
\begin{equation}
\label{Pk}
P_k=\left(\mu-\frac{\sigma^2}{\lambda_k}\right)^+,\ k=1,\ldots,K,
\end{equation} 
where $\mu$ is a dual variable and is given by $$\mu=\frac{\sigma^2e^{2R_{th}\log(2)}}{(\prod_{k=1}^K\lambda_k)^{\frac{1}{K}}}.$$  Therefore, $\mathbf{W}$ can be found using equations \eqref{Pk} $\forall k$ and \eqref{SVD}. 

Now, we can solve problem \eqref{OP45} for $\mathbf{U}$ using the obtained $\mathbf{W}$. The problem in terms of $\mathbf{U}$ can be expressed as follows
\begin{subequations}
\label{OP4}
\begin{eqnarray}
&\displaystyle\min_{\mathbf{U}}& \frac{1}{2}\sum_{k=1}^K\|\mathbf{u}_k\|^2 \\
\label{OP4c}
&\text{s.t.}& \frac{\vert(\mathbf{h}_{I,k}^H\mathbf{\Theta}_2\mathbf{H}_{RI}+\mathbf{h}_{R,k}^H) \mathbf{u}_k\vert^2}{\sum_{i\neq k} \vert (\mathbf{h}_{I,k}^H\mathbf{\Theta}_2\mathbf{H}_{RI}+\mathbf{h}_{R,k}^H)\mathbf{u}_i\vert^2+\sigma^2}\geq \eta_k,
\nonumber\\
&&\ \ \ \ \ \ \  k=1,\ldots, K,
\end{eqnarray}
\end{subequations} 
The optimal solution of problem \eqref{OP4} can be found using either semidefinite programming, second order cone programming, or the uplink-downlink duality. For the sake of simplicity, we adopt the uplink-downlink duality approach. The optimal solution of problem \eqref{OP4} can be found using
\begin{equation}
\label{u}
\mathbf{u}_k=\sqrt{q_k}\bar{\mathbf{u}}_k, k=1,\ldots,K,
\end{equation}
where $q_k$ is the $k$th entry of vector  
\begin{equation}
\label{q}
\mathbf{q}=\sigma^2\mathbf{D}^{-1}\mathbf{1}_{K},
\end{equation}
with $\mathbf{1}_{K}$ being the one vector with length $K$, and $\mathbf{D}$ given by
\begin{equation}
\mathbf{D}_{i,j}=\begin{cases} \frac{\vert \mathbf{h}_{RI,i} \bar{\mathbf{u}}_i\vert^2}{\eta_i}, & i=j;\\
-\vert\mathbf{h}_{RI,i} \bar{\mathbf{u}}_j\vert^2, & i\neq j,
\end{cases}
\end{equation}
and where $\bar{\mathbf{u}}_k$ is given by
\begin{equation}
\label{ubar}
\bar{\mathbf{u}}_k=\frac{(\mathbf{I}_N+\sum_{i=1}^K \frac{\beta_i}{\sigma^2}\mathbf{h}_{RI,i}\mathbf{h}_{RI,i}^H)^{-1}\mathbf{h}_{RI,k}}{\|(\mathbf{I}_N+\sum_{i=1}^K \frac{\beta_i}{\sigma^2}\mathbf{h}_{RI,i}\mathbf{h}_{RI,i}^H)^{-1}\mathbf{h}_{RI,k}\|}, \forall k,
\end{equation}
with $\mathbf{h}_{RI,i}=\mathbf{h}_{I,i}^H\mathbf{\Theta}_2\mathbf{H}_{RI}+\mathbf{h}_{R,i}^H$, and 
\begin{equation}
\label{beta}
\beta_k=\frac{\sigma^2}{(1+\frac{1}{\eta_k})\mathbf{h}_{RI,k}^H(\mathbf{I}_N+\sum_{i=1}^K \frac{\beta_i}{\sigma^2}\mathbf{h}_{RI,i}\mathbf{h}_{RI,i}^H)^{-1}\mathbf{h}_{RI,k}}.
\end{equation}
Thus, first we find the dual variable $\beta_k \ \forall k$ using the fixed point algorithm on \eqref{beta} and then we find $\mathbf{u}_k$  using equations \eqref{u}-\eqref{ubar}. 

\subsubsection{SVD and Zero-Forcing}
Recall that in \eqref{OP1}, we have to find $\mathbf{\Theta}_1$ and $\mathbf{\Theta}_2$ in addition to $\mathbf{W}$ and $\mathbf{U}$. Thus, it is desirable to find simple solutions for $\mathbf{W}$ and $\mathbf{U}$ since they would be used repeatedly in the optimization of $\mathbf{\Theta}_1$ and $\mathbf{\Theta}_2$. Since the above method does not provide a closed-form solution for $\mathbf{U}$ and includes an iterative procedure, we propose another simpler solution next. We propose to design $\mathbf{U}$ to eliminate the interference at the users. This can be obtained using zero-forcing (ZF), where $\mathbf{U}$ is given by
\begin{equation}
\label{UZF}
\mathbf{U}=\mathbf{H}_{RIK}^H(\mathbf{H}_{RIK}\mathbf{H}_{RIK}^H)^{-1}\mathbf{Q}^{\frac{1}{2}}.
\end{equation}
where the $k$th column of $\mathbf{H}_{RIK}$ is $\mathbf{h}_{RI,k}$ and $\mathbf{Q}=\text{diag}[q_1,q_2,\ldots,q_K]$ is the power allocation matrix at the relay. With this design, problem \eqref{OP1} becomes
\begin{subequations}
\label{OP5}
\begin{eqnarray}
&\displaystyle\min_{\mathbf{q}}&   \frac{1}{2}\text{tr}(\mathbf{U}\mathbf{U}^H) \\
\label{OP5c}
&\text{s.t}& q_k\geq \sigma^2\eta_k,\ \ k=1,\ldots, K,
\end{eqnarray}
\end{subequations} 
the solution of which is given by $q_k=\sigma^2\eta_k$. Note that in this method, $\mathbf{W}$ is chosen as in \eqref{SVD}, i.e., using the SVD method.

\subsection{Solution I for $\mathbf{\Theta}_1$ and $\mathbf{\Theta}_2$: SDP Approach}

We aim here to provide solutions for $\mathbf{\Theta}_1$ and $\mathbf{\Theta}_2$ when $\mathbf{W}$ and $\mathbf{U}$ are given. In that case, the problem can be expressed as follows 
\begin{subequations}
\label{OP6}
\begin{eqnarray}
&\displaystyle\min_{\mathbf{\Theta}_1, \mathbf{\Theta}_2}&  \frac{1}{2}\text{tr}(\mathbf{W}\mathbf{W}^H) + \frac{1}{2}\text{tr}(\mathbf{U}\mathbf{U}^H) \\
\label{OP6b}
&\text{s.t.}&   R_{R}\geq 2KR_{th},\\
\label{OP6c}
&& \bar{R}_k\geq 2R_{th},\ \ k=1,\ldots, K,\\
\label{OP6d}
&&\mathbf{\Theta}_1, \mathbf{\Theta}_2 \in \mathcal{F},
\end{eqnarray}
\end{subequations} 
Providing an optimal solution for the above problem is difficult since the relation of $\mathbf{\Theta}_1$ and $\mathbf{W}$ is unknown, $\mathbf{\Theta}_2$ is involved in calculating inverse matrices to find $\mathbf{U}$, and both $\mathbf{\Theta}_1$ and $\mathbf{\Theta}_2$ are coupled through \eqref{OP6c}. To simplify the problem, we first solve it under the assumption that the phase shifts are continuous and that the beamforming matrices $\mathbf{W}$ and $\mathbf{U}$ are given.  We use the alternating optimization method, where we formulate the problem at the $i$th iteration while exploiting the given solutions of the $(i-1)$th iteration. Since the objective function is not a direct function of $\mathbf{\Theta}_1$ and $\mathbf{\Theta}_2$, we formulate the phases optimization problem to maximize the SNRs at the relay and the users, which leads to decreasing the total required power. This is implemented by scaling down the power to achieve the required data rates. In the following, we formulate two separate problems to find $\mathbf{\Theta}_1$ and $\mathbf{\Theta}_2$. For $\mathbf{\Theta}_1$, we formulate the problem at the $i$th iteration as follows

\begin{subequations}
\label{OP66}
\begin{eqnarray}
&\displaystyle\max_{\mathbf{\Theta}_1, \{c_k\}}&  \sum_{k=0}^K c_k\\
\label{OP66b}
&\text{s.t.}&   R_{R}\geq 2KR_{th}+c_0,\\
\nonumber
&& \Scale[0.9]{\gamma_{k,1}\geq\frac{\vert(\mathbf{h}_{I,k}^H\mathbf{\Theta}_1^{(i-1)}\mathbf{H}_{TI}+\mathbf{h}_{T,k}^H) \mathbf{w}_k\vert^2}{\sum_{i\neq k} \vert (\mathbf{h}_{I,k}^H\mathbf{\Theta}_1^{(i-1)}\mathbf{H}_{TI}+\mathbf{h}_{T,k}^H)\mathbf{w}_i\vert^2+\sigma^2}+c_k,}\\
\label{OP66c}
&&\ \ \ \ \ \  k=1,\ldots, K,\\
\label{OP66d}
&&c_k\geq 0, \ k=0,\dots,K,
\end{eqnarray}
\end{subequations}
where $c_k,\ k=0,\ldots,K$ are slack variables and $\mathbf{\Theta}_1^{(i-1)}$ is the result of iteration $i-1$. The expression of $R_R$ given in \eqref{Rr} is a non-convex function of $\mathbf{\Theta}_1$, which prevents applying traditional optimization methods. To optimize $\mathbf{\Theta}_1$, we first find a proper lower approximation of function $R_R$ and then reformulate problem \eqref{OP66} as a semidefinite programming.  Using the relation stating that for any matrices $\mathbf{A}\in \mathbb{C}^{K\times K}$ and $\bar{\bf{A}}\in \mathbb{C}^{K\times K}$  \cite[Page 641]{Boyd} 
\begin{equation}
\label{Apro}
\log_2|\mathbf{A}|\leq \log_2|\bar{\mathbf{A}}|+tr(\bar{\mathbf{A}}^{-1}(\mathbf{A}-\bar{\mathbf{A}})).
\end{equation}
Hence, the function $R_R$ satisfies the following
\begin{align}
R_R&=\log_2\vert \mathbf{I}+\frac{1}{\sigma^2}(\mathbf{H}_{TR}+\mathbf{H}_{IR}\mathbf{\Theta}_1\mathbf{H}_{TI})\mathbf{W}\mathbf{W}^H&\\
&\ \ \ \times(\mathbf{H}_{TR}+\mathbf{H}_{IR}\mathbf{\Theta}_1\mathbf{H}_{TI})^H |& \\
&\leq \log_2|\mathbf{Z}|+\text{tr}(\mathbf{Z}^{-1}(\mathbf{I}+\frac{1}{\sigma^2}(\mathbf{H}_{TR}+\mathbf{H}_{IR}\mathbf{\Theta}_1\mathbf{H}_{TI})&\\
&\ \ \ \times\mathbf{W}\mathbf{W}^H(\mathbf{H}_{TR}+\mathbf{H}_{IR}\mathbf{\Theta}_1\mathbf{H}_{TI})^H -\bf{Z})),&
\end{align}
where 
\begin{multline}
\mathbf{Z}=\mathbf{I}+\frac{1}{\sigma^2}(\mathbf{H}_{TR}+\mathbf{H}_{IR}\mathbf{\Theta}_1^{(i-1)}\mathbf{H}_{TI})\mathbf{W}\bf{W}^H\\
\times(\mathbf{H}_{TR}+\mathbf{H}_{IR}\mathbf{\Theta}_1^{(i-1)}\mathbf{H}_{TI})^H |.
\end{multline}
Let $\mathbf{v}_1=[e^{j\theta_{1,1}},e^{j\theta_{1,2}},\ldots,e^{j\theta_{1,L}}]^H$. Since for any matrices $\mathbf{A}_1$ and $\mathbf{A}_2$, $\text{tr}(\mathbf{\Theta}_1\mathbf{A}_1\mathbf{\Theta}_1\mathbf{A}_2)=\mathbf{v}_1(\mathbf{A}_1\odot \mathbf{A}_2^T)\mathbf{v}_1$ holds, where $\odot$ denotes Hadamard product, so $R_R$ can be upper bounded by
\begin{equation}
\label{R_Rlow}
R_R \leq  F +2 Re \{\mathbf{v}_1^H \mathbf{x} \}+\mathbf{v}_1^H\bar{\mathbf{X}}\mathbf{v}_1,
\end{equation}
where 
\begin{align*}
F&=\log_2|\mathbf{Z}|+\text{tr}(\mathbf{Z}^{-1}\mathbf{I})+\text{tr}(\frac{1}{\sigma^2} \mathbf{H}_{TR} \mathbf{W} \mathbf{Z}^{-1}\mathbf{W}^H\mathbf{H}_{TR}^H)-\text{tr}(\mathbf{I}),&\\
 \bar{\mathbf{X}}&=\frac{1}{\sigma^2}\bf{H}_{TI}\bf{W}\mathbf{Z}^{-1}\mathbf{W}^H\bf{H}_{TI}^H \odot (\mathbf{H}_{IR}^H\bf{H}_{IR})^T,&\\
\mathbf{x}&=\text{diag}(\frac{1}{\sigma^2}\bf{H}_{TI} \bf{W}\bf{Z}^{-1}\bf{W}^H \bf{H}_{TR}^H \bf{H}_{IR}).&
\end{align*}
Define $\mathbf{a}_{k,i}=\text{diag}(\mathbf{h}_{I,k}^H)\mathbf{H}_{TI}\mathbf{w}_i$ and $b_{k,i}=\mathbf{h}_{T,k}^H\mathbf{w}_i$. Then $\gamma_{k,1}$ can be written as follows
\begin{equation}
\gamma_{k,1}=\frac{\bar{\mathbf{v}}_1^H\mathbf{B}_{k,k}\bar{\mathbf{v}}_1+|b_{k,k}|^2}{\sum_{i\neq k} \bar{\mathbf{v}}_1^H\mathbf{B}_{k,i}\bar{\mathbf{v}}_1+\sum_{i\neq k}|b_{k,i}|^2+\sigma^2}.
\end{equation}
where
$\mathbf{B}_{k,i}=  \begin{bmatrix} \mathbf{a}_{k,i}\mathbf{a}_{k,i}^H & \mathbf{a}_{k,i}b_{k,i}^H\\ b_{k,i}\mathbf{a}_{k,i}^H & 0 \end{bmatrix},$ and  $\bar{\mathbf{v}}_1=  \begin{bmatrix} \mathbf{v}_1\\ t\end{bmatrix}.$ By defining  $\mathbf{X}=\begin{bmatrix} \bar{\mathbf{X}} & \mathbf{x}\\\mathbf{x}^H& 0\end{bmatrix}$, problem \eqref{OP66} can be equivalently written as
\begin{subequations}
\label{OP666}
\begin{eqnarray}
&\displaystyle\max_{\mathbf{\Theta}_1, \{c_k\}}&  \sum_{k=0}^K c_k\\
\label{OP666b}
&\text{s.t.}&   \bar{\mathbf{v}}_1^H\mathbf{X}\bar{\mathbf{v}}_1\geq 2KR_{th}+c_0-F,\\
\nonumber
&& \bar{\mathbf{v}}_1^H\mathbf{B}_{k,k}\bar{\mathbf{v}}_1+|b_{k,k}|^2\geq \gamma_{k,1}^{(i-1)}(\sum_{i\neq k}^K \bar{\mathbf{v}}_1^H\mathbf{B}_{k,i}\bar{\mathbf{v}}_1\\
\label{OP666b}
&&+\sum_{i\neq k}|b_{k,i}|^2+\sigma^2)+c_k, \ k=0,\dots,K \\
\label{OP666c}
&&c_k\leq 0; \ k=0,\dots,K, 
\end{eqnarray}
\end{subequations}
where $c_k,\ k=1,\ldots,K$ are slack variables. By defining $\mathbf{V}_1=\bar{\mathbf{v}}_1\bar{\mathbf{v}}_1^H$, problem \eqref{OP666} can be converted to an SDP with rank one constraint as follows
\begin{subequations}
\label{SDP11}
\begin{eqnarray}
&\displaystyle\max_{\mathbf{V}_1, \{c_k\}}&  \sum_{k=0}^K c_k\\
\label{SDP11b}
&\text{s.t.}&   \text{tr}(\mathbf{V}_1\mathbf{X})\geq 2KR_{th}+c_0-F,\\
\nonumber
&& \text{tr}(\mathbf{V}_1\mathbf{B}_{k,k})+|b_{k,k}|^2\geq \gamma_{k,1}^{(i-1)}(\sum_{i\neq k}^K \text{tr}(\mathbf{V}_1\mathbf{B}_{k,i})\\
\label{SDP11b}
&&+\sum_{i\neq k}|b_{k,i}|^2+\sigma^2)+c_k, \ k=0,\dots,K \\
\label{SDP11c}
&&c_k\leq 0; \ k=0,\dots,K,\\
\label{SDP11d}
&& \mathbf{V}_1\succeq 0, \text{Rank}(\mathbf{V}_1)=1
\end{eqnarray}
\end{subequations}
To solve  problem \eqref{SDP11}, we first relax the rank-one constraint and then solve the problem using the CVX solver \cite{cvx}. The resulting $\mathbf{V}_1$ is not guaranteed to be with rank one, so we use the randomization method to obtain a rank one solution. The randomization method is explained in \cite{wu2018intelligent}. 

Similarly, we can find $\mathbf{\Theta}_2$ by solving the following problem
\begin{subequations}
\label{OP65}
\begin{eqnarray}
&\displaystyle\max_{\mathbf{\Theta}_2, \{c_k\}}&  \sum_{k=1}^K c_k\\
\nonumber
&& \Scale[0.9]{\gamma_{k,2}\geq\frac{\vert(\mathbf{h}_{I,k}^H\mathbf{\Theta}_2^{(i-1)}\mathbf{H}_{RI}+\mathbf{h}_{R,k}^H) \mathbf{u}_k\vert^2}{\sum_{i\neq k} \vert (\mathbf{h}_{I,k}^H\mathbf{\Theta}_2^{(i-1)}\mathbf{H}_{RI}+\mathbf{h}_{R,k}^H)\mathbf{u}_i\vert^2+\sigma^2}+c_k,}\\
\label{OP65b}
&&\ \ k=1,\ldots, K,\\
\label{OP65c}
&&c_k\geq 0; \ k=1,\dots,K 
\end{eqnarray}
\end{subequations}
Following similar steps to those used to obtain  $\mathbf{\Theta}_1$, we first convert the problem into an SDP by defining that $\mathbf{V}_2=\bar{\mathbf{v}}_2\bar{\mathbf{v}}_2^H$, $\bar{\mathbf{v}}_2=\begin{bmatrix} \mathbf{v}_2\\ t\end{bmatrix},$ $\mathbf{v}_2=[e^{j\theta_{2,1}},e^{j\theta_{2,2}},\ldots,e^{j\theta_{2,L}}]^H$, then the matrix $\mathbf{V}_2$ can be obtained by solving the following problem after relaxing the rank one constraint 
\begin{subequations}
\label{SDP12}
\begin{eqnarray}
&\displaystyle\max_{\mathbf{V}_2, \{c_k\}}&  \sum_{k=0}^K c_k\\
\nonumber
&& \text{tr}(\mathbf{V}_2\mathbf{G}_{k,k})+|g_{k,k}|^2\geq \gamma_{k,2}^{(i-1)}(\sum_{i\neq k}^K \text{tr}(\mathbf{V}_2\mathbf{G}_{k,i})\\
\label{SDP12b}
&&+\sum_{i\neq k}|g_{k,i}|^2+\sigma^2)+c_k, \ k=0,\dots,K \\
\label{SDP12c}
&&c_k\leq 0; \ k=0,\dots,K,\\
\label{SDP12d}
&& \mathbf{V}_2\succeq 0, \text{Rank}(\mathbf{V}_2)=1
\end{eqnarray}
\end{subequations}
where $\mathbf{G}_{k,i}= \begin{bmatrix} \mathbf{z}_{k,i}\mathbf{z}_{k,i}^H & \mathbf{z}_{k,i}g_{k,i}^H\\ g_{k,i}\mathbf{z}_{k,i}^H & 0 \end{bmatrix},$ $\mathbf{z}_{k,i}=\text{diag}(\mathbf{h}_{I,k}^H)\mathbf{H}_{RI}\mathbf{u}_i$ $g_{k,i}=\mathbf{h}_{R,k}^H\mathbf{u}_i$. If the resulting $\mathbf{V}_2$ is not of rank one, we use the randomization method to get a rank-one solution.

After finding the solutions $\bar{\mathbf{v}}_1$ and $\bar{\mathbf{v}}_2$, we can find the phase shifts using the following expressions
\begin{equation}
\label{thet11}
\theta_{1,\ell}=\exp\bigg(j \arg\bigg(\frac{\bar{v}_{1,\ell}^H}{\bar{v}_{1,L+1}^H}\bigg)\bigg)
\end{equation}
\begin{equation}
\label{thet22}
\theta_{2,\ell}=\exp\bigg(j \arg\bigg(\frac{\bar{v}_{2,\ell}^H}{\bar{v}_{2,L+1}^H}\bigg)\bigg)
\end{equation}
Overall, the steps for finding $\mathbf{W}$, $\mathbf{U}$, $\mathbf{\Theta}_1$, and $\mathbf{\Theta}_2$ are given in Algorithm \ref{algor1}.

\begin{algorithm}[t]
\SetAlgoLined
 \label{algor1}
 Initialize $\mathbf{\Theta}_1$ and $\mathbf{\Theta}_2$\;
 \Repeat {$\vert \text{tr}(\mathbf{W}\mathbf{W}^H)^{(i)}+ \text{tr}(\mathbf{U}\mathbf{U}^H)^{(i)}-\big(\text{tr}(\mathbf{W}\mathbf{W}^H)^{(i-1)}+\text{tr}(\mathbf{U}\mathbf{U}^H)^{(i-1)}\big)\vert\leq \epsilon$} {
 Find $\mathbf{W}$ using equations \eqref{Pk} $\forall k$ and \eqref{SVD}\;
 Find $\mathbf{U}$ using equations \eqref{u}-\eqref{ubar}\;
 Find $\mathbf{\Theta}_1$ by solving problem \eqref{SDP11} and then using \eqref{thet11}\;
 Find $\mathbf{\Theta}_2$ by solving problem \eqref{SDP12} and then using \eqref{thet22}\;
  } 
 Find $\mathbf{W}$ using equations \eqref{Pk} $\forall k$ and \eqref{SVD}\;
 Find $\mathbf{U}$ using equations \eqref{u}-\eqref{ubar}\;
 \caption{Finding $\mathbf{W}$, $\mathbf{U}$, $\mathbf{\Theta}_1$, and $\mathbf{\Theta}_2$ using SVD, uplink-downlink duality, and SDP.}
 \end{algorithm}
 
 It is easy to prove the convergence of Algorithm \ref{algor1} since at each iteration the objective function either decreases or stays fixed, and it is bounded from below.  

There are two disadvantage in solving the problem of $\mathbf{\Theta}_1$, and $\mathbf{\Theta}_2$ using SDP with relaxation. The first one is that SDP does not guarantee the optimal solution due to the relaxation. The second one is that the computational complexity of SDP is of order $O((L+1)^6)$, which means the complexity increases dramatically with increasing the number of phase shifts. Due to the fact that RIS would be competitive only if the number of reflecting elements is large \cite{sonR25}, using SDP to find a solution for the phase shifts would be computationally prohibitive. Therefore, we propose another simpler solution that has much lower complexity than the SDP approach with negligible performance loss. 
\subsection{Solution II for $\mathbf{\Theta}_1$ and $\mathbf{\Theta}_2$: Maximizing SNR of the Weakest Hop}
In this approach, we also aim to find solutions for $\mathbf{\Theta}_1$ and $\mathbf{\Theta}_2$ to maximize the SNR at the relay and users. To find a low-complexity solution, we formulate the optimization problem in a simpler way. For $\mathbf{\Theta}_1$, we design the phase shifts to maximize the minimum rate of both hops (i.e., either the SNR at the relay or the SINR at the users). In other words, we formulate two optimization problems: One for maximizing the rate at the relay, and the other for maximizing the sum of the power of received desired signal at the users. We then solve both problems and select the one that leads to lower total power. To maximize the rate at the relay, we formulate a problem that maximizes a surrogate function of $R_R$ that is given by \eqref{R_Rlow}. So the problem is given as follows
\begin{subequations}
\label{Re1}
\begin{eqnarray}
\label{Re1a}
&\displaystyle\max_{\mathbf{v}_1}& F +2 Re \{\mathbf{v}_1^H \mathbf{x} \}+\mathbf{v}_1^H\bar{\mathbf{X}}\mathbf{v}_1,\\
\label{Re1b}
&\text{s.t.}& |v_{1,i}|=1,\ i=1,\ldots,L
\end{eqnarray}
\end{subequations}
Problem \eqref{Re1} is difficult because of the non-convex constraint. To simplify it, we propose to use an approximation for the third term in the objective function, which is given as follows \cite{7547360}
\begin{multline}
\label{Ineq}
\mathbf{v}_1^H\bar{\mathbf{X}}\mathbf{v}_1\leq \mathbf{v}_1^H\lambda_{max}(\bar{\mathbf{X}})\mathbf{I}\mathbf{v}_1+2\text{Re}\{\mathbf{v}_1^H(\bar{\mathbf{X}}-\lambda_{max}(\bar{\mathbf{X}})\mathbf{I})\hat{\mathbf{v}}_1\}\\
+\hat{\mathbf{v}}_1^H(\lambda_{max}(\mathbf{X})\mathbf{I}-\mathbf{X})\hat{\mathbf{v}}_1.
\end{multline}
 where  $\hat{\mathbf{v}}_1$ is any feasible point and $\lambda_{max}(\mathbf{X})$ is the maximum eigenvalue of $\mathbf{X}$. In \eqref{Ineq}, the equality holds when $\mathbf{v}_1=\hat{\mathbf{v}}_1$. Using the approximation in \eqref{Ineq} and since $\mathbf{v}_1^H\lambda_{max}(\bar{\mathbf{X}})\mathbf{I}\mathbf{v}_1=L\lambda_{max}(\bar{\mathbf{X}})$, problem \eqref{Re1} can be approximated as follows 
\begin{subequations}
\label{Re2}
\begin{eqnarray}
\label{Re2a}
&\displaystyle\max_{\mathbf{v}_1}& \text{Re}\{\mathbf{q}^H\mathbf{v}_1\},\\
\label{Re2b}
&\text{s.t.}& |v_{1,i}|=1,\ i=1,\ldots,L
\end{eqnarray}
\end{subequations}
where $\mathbf{q}=\mathbf{x}+(\bar{\mathbf{X}}-\lambda_{max}(\bar{\mathbf{X}})\mathbf{I})\hat{\mathbf{v}}_1$. The solution of problem \eqref{Re2} is given by 
\begin{equation}
\label{v1}
\mathbf{v}_1=[e^{j\arg(q_1)}, e^{j\arg(q_2)},\ldots,e^{j\arg(q_L)}]^T.
\end{equation}

The expression provided in \eqref{v1} is the solution that maximizes the rate at the relay in the first phase. However, when the BS-relay link is sufficiently strong, it may be better to design $\mathbf{\Theta}_1$ to maximize the rate at the users. To design $\mathbf{\Theta}_1$ using low complexity approach, we formulate a problem to maximize the sum of the received desired power at the users under given beamforming matrices. In particular, we formulate the problem as follows
 
\begin{subequations}
\label{Re3}
\begin{eqnarray}
\label{Re3a}
&\displaystyle\max_{\bar{\mathbf{v}}_1}& \sum_{k=1}^K(\bar{\mathbf{v}}_1^H\mathbf{B}_{k,k}\bar{\mathbf{v}}_1),\\
\label{Re3b}
&\text{s.t.}& |\bar{v}_{1,i}|=1,\ i=1,\ldots,L
\end{eqnarray}
\end{subequations} 
To solve problem \eqref{Re3}, we use the fixed point iteration that guarantees a local optimal solution \cite{8855810}. This approach is an iterative algorithm where in the $i$th iteration, the value of $\bar{\mathbf{v}}_1$ is updated as follows
\begin{equation}
\label{v1bar}
\bar{\mathbf{v}}_1^{i}= \text{unt}\bigg(\sum_{k=1}^K\mathbf{B}_{k,k}\bar{\mathbf{v}}_1^{i-1}\bigg),
\end{equation} 
where $\text{unt}(\mathbf{a})$ is the vector whose elements are $\frac{a_1}{|a_1|},\frac{a_2}{|a_2|},\ldots, \frac{a_{L+1}}{|a_{L+1}|}$. Then, the phase shifts of $\mathbf{\Theta}_1$ can be found by 
\begin{equation}
\label{tht1ell}
\theta_{1,l}=\exp\bigg(j \arg\bigg(\frac{\bar{v}_{1,l}^H}{\bar{v}_{1,L+1}^H}\bigg)\bigg)
\end{equation}

For updating $\mathbf{\Theta_2}$, we use the same approach used to find $\mathbf{\Theta_1}$. Since $R_R$ is not a function of $\mathbf{\Theta_2}$, we only need to maximize the received power of the desired signal coming from the relay and this can be formulated as follows
\begin{subequations}
\label{Re4}
\begin{eqnarray}
\label{Re4a}
&\displaystyle\max_{\bar{\mathbf{v}}_2}& \sum_{k=1}^K(\bar{\mathbf{v}}_2^H\mathbf{G}_{k,k}\bar{\mathbf{v}}_2),\\
\label{Re4b}
&\text{s.t.}& |\bar{v}_{2,i}|=1,\ i=1,\ldots,L.
\end{eqnarray}
\end{subequations} 
 Hence, the solution for $\mathbf{\Theta_2}$ in the $i$th iteration is given by 
\begin{equation}
\label{v2bar}
\bar{\mathbf{v}}_2^{i}= \text{unt}\bigg(\sum_{k=1}^K\mathbf{G}_{k,k}\bar{\mathbf{v}}_2^{i-1}\bigg).
\end{equation}
Similarly, the phase shifts of $\mathbf{\Theta}_2$ are given by 
\begin{equation}
\label{tht2}
\theta_{2,l}=\exp\bigg(j \arg(\frac{\bar{v}_{2,l}^H}{\bar{v}_{2,L+1}^H}\bigg)\bigg)
\end{equation} 
Algorithm \ref{algor2} provides the steps of the second approach in details.
\begin{algorithm}[t]
\SetAlgoLined
 \label{algor2}
 Initialize $\mathbf{\Theta}_1$ and $\mathbf{\Theta}_2$\;
 \Repeat {$\vert \text{tr}(\mathbf{W}\mathbf{W}^H)^{(i)}+\text{tr}(\mathbf{U}\mathbf{U}^H)^{(i)}-\big(\text{tr}(\mathbf{W}\mathbf{W}^H)^{(i-1)}+\text{tr}(\mathbf{U}\mathbf{U}^H)^{(i-1)}\big)\vert\leq \epsilon_1$}{
 Find $\mathbf{W}$ using equations \eqref{Pk} $\forall k$ and \eqref{SVD}\;
 Find $\mathbf{U}$ using equations \eqref{u}-\eqref{ubar}\;
 Find $\mathbf{\Theta}_1$ using the expression \eqref{v1} and \eqref{tht1ell} and assign it to $\mathbf{\Theta}_{1,1}$\;
 Set $j=1$\;
 \Repeat {$\vert \bar{\mathbf{v}}_1^{(j)}-\bar{\mathbf{v}}_1^{(j-1)}\vert\leq \epsilon_2$}{
 Update $\bar{\mathbf{v}}_1^{(j)}$ using \eqref{v1bar}\;
 Set $j=j+1$\;
 }
 Find $\mathbf{\Theta}_1$ using \eqref{tht1ell} and assign it to $\mathbf{\Theta}_{1,2}$\;
 Select $\mathbf{\Theta}_{1,1}$ or $\mathbf{\Theta}_{1,2}$ that produces lower value of \eqref{OP1a}\ and assign it to $\mathbf{\Theta}_{1}$\;
 Set $j=1$\;
 \Repeat {$\vert \bar{\mathbf{v}}_2^{(j)}-\bar{\mathbf{v}}_2^{(j-1)}\vert\leq \epsilon_2$}{
 Update $\bar{\mathbf{v}}_2$ using \eqref{v2bar}\;
 Set $j=j+1$\;
 }
 Find $\mathbf{\Theta}_2$ using \eqref{tht2} \;
 }
 Find $\mathbf{W}$ using equations \eqref{Pk} $\forall k$ and \eqref{SVD}\;
 Find $\mathbf{U}$ using equations \eqref{u}-\eqref{ubar}\;
 \caption{Finding $\mathbf{W}$, $\mathbf{U}$, $\mathbf{\Theta}_1$, and $\mathbf{\Theta}_2$ using SVD, uplink-downlink duality, and Approach 2.}
 \end{algorithm}

The computational complexity of Algorithm \ref{algor2} is much lower than that of Algorithm \ref{algor1}. This is because in  Algorithm \ref{algor2} we update the phase shifts in each iteration using a closed-form expressions, while in Algorithm \ref{algor1}, we implement two SDP problems that entail a computation complexity of $O(2(L+1)^6)$. 

Next, we formulate the problem for minimizing the total transmit power subject to QoS constraints in the full-duplex mode.

\section{Problem Formulation And Proposed Algorithms: Full-Duplex Mode} \label{Sec:Full}
In this section, we study the problem of minimizing the transmit power of the proposed system under full-duplex relay operation. Under this assumption, the ptransmission takes place over one phase only (instead of two), where we only need to optimize $L$ phase shifts collected in a $\mathbf{\Theta}\in \mathbb{C}^{L\times L}$ (instead of $2L$ phases in two matrices $\mathbf{\Theta}_1$ and $\mathbf{\Theta}_2$). 

The optimization problem can be written now as follows
\begin{subequations}
\label{OP7}
\begin{eqnarray}
&\displaystyle\min_{\mathbf{W}, \mathbf{\Theta},\mathbf{U}}& \sum_{k=1}^K\|\mathbf{w}_k\|^2+\sum_{k=1}^K\|\mathbf{u}_k\|^2 \\
\label{OP7b}
&\text{s.t.}&   R_{R}^{FD}\geq KR_{th},\\
\label{OP7c}
&& R_{k}^{FD}\geq R_{th},\ \ k=1,\ldots, K,\\
\label{OP7d}
&&\mathbf{\Theta} \in \mathcal{F},
\end{eqnarray}
\end{subequations} 
Problem (\ref{OP7}) is different from problem (\ref{OP1}) since there is only one phases matrix $\mathbf{\Theta}$ to be found and the constraints in (\ref{OP7c}) are different form those in (\ref{OP1c}). To solve such problem, we first should note that the problem is similar to a quadratically-constrained-quadratic program with an additional log determinant constraint (\ref{OP7b}). The main difficulty in this problem is that the matrix $\mathbf{W}$ affects the rate at the relay and the users. Since we assumed that the relay is close to the RIS (to strengthen the relay-RIS channels) and the BS is too far from RIS ($> 200$ m) (and that the direct relay-user channels are much stronger than the direct BS-user channels), it holds with probability one that $\sum_{i\neq k}^K \vert \mathbf{h}_{IR,k}\mathbf{u}_i\vert^2 \gg \sum_{i=1}^K\vert \mathbf{h}_{TI,k}\mathbf{w}_i\vert^2$. This means that the interference at the user $k$  from the BS is negligible compared to the interference from the relay. We propose a solution that relies on this fact and ignore the effect of the BS interference at the users. In other words, in our solution, we ignore the term $\sum_{i=1}^K\vert \mathbf{h}_{TI,k}\mathbf{w}_i\vert^2$ in equation \eqref{Rkfd} at the user $k$ $\forall K$. Under this assumption, the coupling between variables $\mathbf{U}$ and $\mathbf{W}$ in problem \eqref{OP7} is eliminated,  and hence problem \eqref{OP7} can be divided (approximately) into two independent problems.  
\subsection{Optimizing $\mathbf{W}$ and $\mathbf{U}$}
The problem of optimizing with respect to $\mathbf{W}$ can be formulated similar to problem \eqref{OP2} while dropping the half-duplex constraint and can be also solved optimally using SVD and water filling. Therefore, the solution of $\mathbf{W}$ is given by \eqref{SVD}, where $P_k$ is given by \eqref{Pk}, and $\mu=\frac{\sigma^2e^{R_{th}\log(2)}}{(\prod_{k=1}^K\lambda_k)^{\frac{1}{K}}}.$ 

After finding $\mathbf{W}$, problem \eqref{OP7} can now be solved for $\mathbf{U}$ using the uplink-downlink duality approach.  The solution for $\mathbf{U}$ in the full-duplex system is similar to the solution given in equations \eqref{u}-\eqref{beta}, with $\sigma^2$ replaced by $\sum_{i=1}^K\vert \mathbf{h}_{TI,k}\mathbf{w}_i\vert^2+\sigma^2$ and with setting $\eta_k=2^{R_{th}}-1$.

To find a simpler solution for $\mathbf{U}$, we can use zero-forcing. In this approach, $\mathbf{U}$ is given by \eqref{UZF}, where the diagonal matrix $\mathbf{Q}$ is given by $\mathbf{Q}=\text{diag}[q_1,q_2,\ldots,q_K]$ and $q_k=(2^{R_{th}}-1)(\sum_{i=1}^K\vert \mathbf{h}_{TI,k}\mathbf{w}_i\vert^2+\sigma^2)$.
\subsection{Solution I for $\mathbf{\Theta}$: SDP Approach}
Now, for the given $\mathbf{W}$ and $\mathbf{U}$, we can optimize problem \eqref{OP7} in terms of $\mathbf{\Theta}$. Similar to the proposed solutions in half-duplex scheme, we propose here two solutions to find $\mathbf{\Theta}$: One based on SDP, and the other based on maximizing the sum of the received desired signals.  

We formulate the problem of finding $\mathbf{\Theta}$ as a received SNR is maximization problem. This allows the required power to be scaled down as much as possible when we solve the problem for $\mathbf{W}$ and $\mathbf{U}$ for a given $\mathbf{\Theta}$. Then the problem is solved iteratively by alternating between finding $\mathbf{\Theta}$ and $(\mathbf{W},\mathbf{U})$. The SNR maximization problem which can be solved to find $\mathbf{\Theta}$ can be written as 
\begin{subequations}
\label{SDP3}
\begin{eqnarray}
&\displaystyle\max_{\mathbf{V}, \{c_k\}}&  \sum_{k=0}^K c_k\\
\label{SDP3b}
&\text{s.t.}&   \text{tr}(\mathbf{V}\mathbf{X})\geq KR_{th}+c_0,\\
\nonumber
&& \text{tr}(\mathbf{V}\mathbf{G}_{k,k})+|g_{k,k}|^2\geq \\
\nonumber
&&c_k+\gamma_{k}^{(i-1)}\big(\sum_{i\neq k}^K \big(\text{tr}(\mathbf{V}\mathbf{G}_{k,i})+|g_{k,i}|^2\big)\\
\label{SDP3c}
&&+\sum_{i=1}^K \big(\text{tr}(\mathbf{V}\mathbf{B}_{k,i})+|b_{k,i}|^2\big)+\sigma^2\big), \ k=0,\dots,K \\
\label{SDP3d}
&&c_k\geq 0; \ k=0,\dots,K,\\
\label{SDP3e}
&& \mathbf{V}\succeq 0, \text{Rank}(\mathbf{V})=1
\end{eqnarray}
\end{subequations}
where $\mathbf{V}=\bar{\mathbf{v}}\bar{\mathbf{v}}^H$, $\bar{\mathbf{v}}=[\mathbf{v}\ t]^T$, $\mathbf{v}=[e^{j\theta_{1}},e^{j\theta_{2}},\ldots,e^{j\theta_{L}}]^H$. $\mathbf{G}_{k,i}$, $\mathbf{B}_{k,i}$, $g_{k,i}$, and $b_{k,i}$ are defined earlier in Section \ref{Sec:Form}. It is important to note that constraint \eqref{SDP3b} is an approximation of constraint \eqref{OP7b} which can be obtained using  \eqref{Apro} and following the steps thereafter. By dropping the rank constraint, problem \eqref{SDP3} can be solved using the CVX solver \cite{cvx}. If the resulting matrix $\mathbf{V}$ is not with a rank one, the randomization method can be used to obtain a rank one solution.
\subsection{Solution II for $\mathbf{\Theta}$: Maximizing SNR of the Weakest Hop} \label{Sec:ApII}
As mentioned before, the complexity of the SDP approach is quite high and increases dramatically with increasing the number of trace constraints. In other words, for a high number of phase shifts, the SDP approach is prohibitively complex. Therefore, we propose another low-complexity solution that achieves similar performance as the SDP approach. The idea is to design $\mathbf{\Theta}$ either to maximize the rate at the relay or at the users. Then the solution that minimizes the total power is selected. For maximizing the achievable rate at the relay, the problem is similar to \eqref{Re1}, and the solution is given by
\begin{equation}
\label{vv}
\mathbf{v}=[e^{j\arg(q_1)}, e^{j\arg(q_2)},\ldots,e^{j\arg(q_L)}]^T.
\end{equation}
To maximize the rate at the users, the problem is similar to \eqref{Re4}, and the solution is given by
\begin{equation}
\label{vv2}
\bar{\mathbf{v}}^{i}= \text{unt}\bigg(\sum_{k=1}^K\mathbf{G}_{k,k}\bar{\mathbf{v}}^{i-1}\bigg).
\end{equation}
Then, the phase shifts can be found using the following
\begin{equation}
\label{tht}
\theta_{l}=\exp\bigg(j \arg\bigg(\frac{\bar{v}_{\ell}^H}{\bar{v}_{L+1}^H}\bigg)\bigg)
\end{equation}
Algorithm \ref{algor3} provides the steps needed for finding the beamforming matrices $\mathbf{W}$ and $\mathbf{U}$ and the phase shifts matrix $\mathbf{\Theta}$ using the second approach proposed in Section \ref{Sec:ApII}.
\begin{algorithm}[t]
\SetAlgoLined
 \label{algor3}
 Initialize $\mathbf{\Theta}$\;
 \Repeat {$\vert \text{tr}(\mathbf{W}\mathbf{W}^H)^{(i)}+\text{tr}(\mathbf{U}\mathbf{U}^H)^{(i)}-\big(\text{tr}(\mathbf{W}\mathbf{W}^H)^{(i-1)}+\text{tr}(\mathbf{U}\mathbf{U}^H)^{(i-1)}\big)\vert\leq \epsilon_1$}
 {
 Find $\mathbf{W}$ using equations \eqref{Pk} $\forall k$ and \eqref{SVD}\;
 Find $\mathbf{U}$ using equations \eqref{u}-\eqref{ubar}\;
 Find $\mathbf{\Theta}$ by using expression \eqref{vv} and \eqref{tht} and assign it to $\mathbf{\Theta}^{(1)}$\;
 \Repeat {$\vert \bar{\mathbf{v}}_1^{(j)}-\bar{\mathbf{v}}_1^{(j-1)}\vert\leq \epsilon_2$}{
 Update $\bar{\mathbf{v}}$ using \eqref{vv2}\;
 }
 Find $\mathbf{\Theta}$ using \eqref{tht} and assign it to $\mathbf{\Theta}^{(2)}$\;
 Select $\mathbf{\Theta}^{(1)}$ or $\mathbf{\Theta}^{(2)}$ that minimizes \eqref{OP1a} and assign it to $\mathbf{\Theta}$\;
 }
 Find $\mathbf{W}$ using equations \eqref{Pk} $\forall k$ and \eqref{SVD}\;
 Find $\mathbf{U}$ using equations \eqref{u}-\eqref{ubar}\;
 \caption{Full-duplex: Finding $\mathbf{W}$, $\mathbf{U}$, and $\mathbf{\Theta}$ using SVD, uplink-downlink duality, and Approach II.}
 \end{algorithm}

\section{Discrete Phases and Benchmarks} \label{Sec:Spec}
In this section, we discuss some scenarios which will be used for comparison with the solutions presented above. First, we discuss the case where the phase shifts applied by the RIS are taken from a discrete set. Then we will discuss the system without an RIS and the system without a relay as benchmarks. 
\subsection{Case 1: Discrete Phase-Shift at RIS}
We provide two solutions for the case where the phase shifts at RIS are discrete. The first solution is based on exploiting the solution of continuous phase shifts (whether the solution of SDP or the one provided in Algorithm \ref{algor2}) and round each phase shift to the closest feasible one. 
The second solution is not based on alternating optimization of $\mathbf{W}$, $\mathbf{U}$, and $\mathbf{\Theta}_1$ and $\mathbf{\Theta}_1$ in the half-duplex case (or $\mathbf{\Theta}$ in the full-duplex case). It is instead based on a successive refinement over the space of discrete phase shifts, wherein at each refinement stage both $\mathbf{W}$ and $\mathbf{U}$ are updated.  Authors of \cite{cui2014coding, kaina2014shaping} show that the practical number of bits of phase shifts is limited (e.g., $b=1$ or $b=2$), leading to $2^b$ phase shifts. This means that the search space of each phase shift is small if the other phase shifts are fixed. Therefore, for the half-duplex case (or full-duplex case),  we adopt the an approach that fixes $2L-1$ (in half-duplex case) or $L-1$ (in full-duplex case) phase shifts and finds the optimal solution for only one phase shift at a time using a one-dimensional search. For each step in this search, new values for $\mathbf{W}$ and $\mathbf{U}$ are found using SVD and uplink-downlink duality approaches, respectively. We repeat this until we go over all the phase shifts in $\mathbf{\Theta}_1$ and $\mathbf{\Theta}_2$ (or $\mathbf{\Theta}$ in full-duplex case). This approach is called successive refinement. This means that new values for $\mathbf{W}$ and $\mathbf{U}$ will be calculated $2L\times 2^b$ times (or $L\times 2^b$ times for the full-duplex case).      


\subsection{Case 2: Relay-Assisted Multi-user System (Without RIS)} 
As a benchmark, we would like to compare the performance of the proposed system with one that does not have an RIS. To evaluate the performance in this case, we solve optimization problems \eqref{OP1} and \eqref{OP7} when the RIS is absent. The problem in this case is solved using the same procedures, but we set $\mathbf{\Theta}_1=\mathbf{0}$ and $\mathbf{\Theta}_2=\mathbf{0}$ in the half-duplex mode and $\mathbf{\Theta}=\mathbf{0}$ in the full-duplex mode. This benchmark would help us to quantify the power reduction achieved due to the RIS.

\subsection{Case 3: RIS-Assisted Multi-user System (Without Relay)}
As another benchmark, we would like to compare with a system that does not have a relay. This would help us quantify the power reduction achieved by the relay only. To evaluate the performance in this case, we  solve problems \eqref{OP1} and \eqref{OP7} when the relay is absent. In this case, there is no point in imposing the half-duplex constraint. Hence, we solve the problem only in the full-duplex mode, while setting $\mathbf{U}=\mathbf{0}$ (no relay).  The problem of optimizing $\mathbf{W}$ becomes the same as problem \eqref{OP4} and can be solved using uplink-downlink duality, while $\mathbf{\Theta}$ can be found by  either the SDP approach or similar to the approach provided in Algorithm \ref{algor2}.

\section{Simulation Results}\label{Sec:Sim}
\begin{figure}[!t]
\centering
\includegraphics[width=3in]{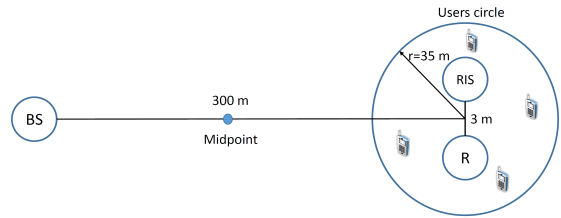}
\caption{Simulation setup.}
\label{setup}
\end{figure}
\begin{figure}[!t]
\centering
\includegraphics[width=3in]{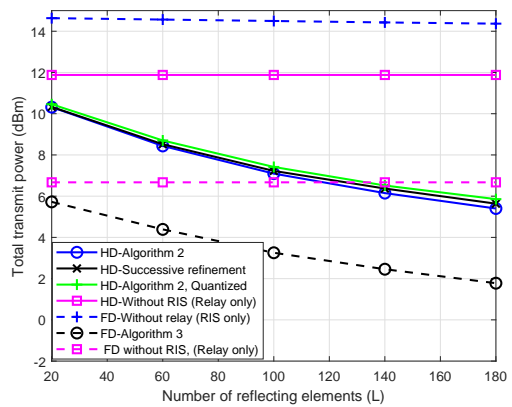}
\caption{The required total power versus the required number of reflecting elements when $R_{th}=2$, $K=4$ $N=5$, $M=5$, $L=50$, the relay and the RIS are in the mid distance between the BS and the users center.}
\label{P_L2}
\end{figure}

In the following simulations, we assume that the antennas at the BS and relay form a uniform linear array (ULA), while the reflecting elements at the RIS form a uniform planner array (UPA). The distances between the antennas and reflecting elements are calculated based on this deployment. As shown in Fig. \ref{setup}, the users are randomly distributed inside a circle of radius $r=35$ m, whose center is $300$ m away from the BS.  The relay and the RIS are assumed to be at the users' circle center except in Fig. \ref{P_L2}, where we assume the relay and the RIS are at the midpoint between the BS and the users circle center. In all simulations, unless otherwise noted, we assume that $M=5$, $N=5$, $K=4$, and $L=50$. The number of bits $b$ is assumed to be $b=2$, so each reflecting element can have $2^b$ different phases, which are $\mathcal{F}=[0, \Delta \theta, \ldots, (2^b-1)\Delta \theta]$, where $\Delta \theta=\frac{2 \pi}{2^b}$. Each point in the simulation figures is an average of 100 realization, where at each we distribute the users inside the circle randomly. 
\begin{figure}[!t]
\centering
\includegraphics[width=3in]{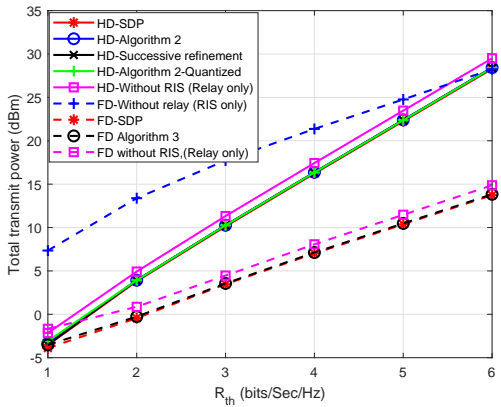}
\caption{The required total power versus the required $R_{th}$ with different number of users $K$ when $N=5$, $M=5$, $L=50$.}
\label{P_Rth}
\end{figure}

We evaluate the impact of changing the required rates at the users ($R_{th}$), the number of users, and the number of reflecting elements on the required total power of the system. We assume that the channels between BS-relay, BS-RIS, and relay-RIS are all modelled as a Rician fading channels, where the LoS is available. We also assume that the channels of the three nodes (BS, relay, and RIS) and a  user is a Rayleigh fading channel, where the LoS is not available. The channel attenuation coefficient between any two points is given by $\beta=\frac{C}{d^{\alpha}}$, where $C=10^{\frac{G_t+G_r-35.95}{10}}$, $\alpha=2.2$ if the LoS is available, and $C=10^{\frac{G_t+G_r-33.95}{10}}$, $\alpha=3.67$ if the LoS is not available, where $G_t=5$ dBi and $G_r=0$ dBi are the antenna gains in dBi at the transmitter and the receiver \cite{NadeemR4}.

Fig. \ref{P_Rth} shows how the total power behaves as a function of $R_{th}$ with the different approaches and different systems. It shows that the full-duplex proposed system (relay+RIS) performs better than all the other systems over all values of $R_{th}$. In the full-duplex mode, the figure shows that the average contribution of power reduction over all values of $R_{th}$ is $1.3$ dBm, while the contribution of full-duplex relay is $13.76$ dBm. However, RIS performs better than the half-duplex mode when the $R_{th}$ value is greater than 6. The figure also shows that the full-duplex and half-duplex systems roughly provide the same performance when $R_{th}$ is small (e.g., $R_{th}= 1$), while at the high values of $R_{th}$, the system with an RIS and without a relay starts to perform better than the half-duplex system. From the algorithms' perspective, although the figure shows the superiority of the SDP approach in both the full and half-duplex cases, the performance loss of Algorithm \ref{algor2} or Algorithm \ref{algor3} compared to the SDP approach is negligible.

\begin{figure}[!t]
\centering
\includegraphics[width=3in]{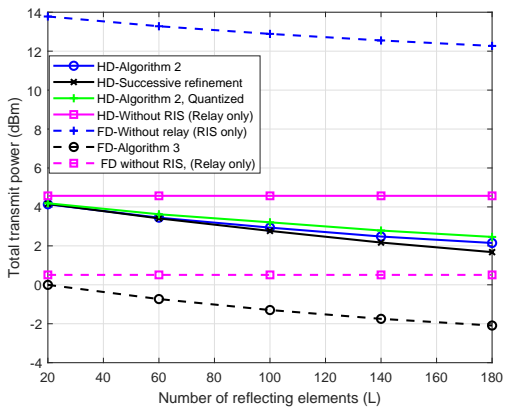}
\caption{The required total power versus the required number of reflecting elements when $R_{th}=2$, $K=4$ $N=5$, $M=5$, the relay and the RIS are at the users center.}
\label{P_L1}
\end{figure}

\begin{figure}[!t]
\centering
\includegraphics[width=3in]{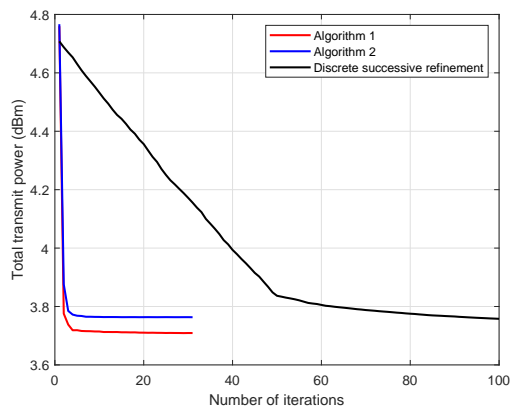}
\caption{The required total power versus the required number of iterations, $L=50$, and $R_{th}=2$.}
\label{P_Itr}
\end{figure}

Fig. \ref{P_L1} shows the effect of increasing the number of reflecting elements at the RIS on the total required power. The figure shows that increasing the number of reflecting elements decreases the required power at BS and the relay. In the half-duplex mode, it can be seen that as the number of reflecting elements increases the successive refinement approach starts to perform better than Algorithm \ref{algor2}. However, since the number of required iterations in the successive refinement approach depends on the number of reflecting elements, its complexity becomes significantly higher compared to Algorithm \ref{algor2}. In particular, when $L=180$, the successive refinement approach solves the problem of finding $\mathbf{W}$ and $\mathbf{U}$ $2^2\times L=720$ times, which is much longer than what algorithm \ref{algor2} needs to converge (cf. Fig. \ref{P_Itr}).
 The figure also shows that increasing the number of reflecting elements from 20 to 180 would reduce the required power by 1.5117 dBm if there is no relay (RIS-only) and 2.0882 dBm if there is a full-duplex relay with 5 antennas (the proposed model). This means that increasing the number of reflecting elements is more impactful in the presence of the relay than in its absence. To highlight this point further, we examine the effect of increasing $L$ when the channel of the second hop is weak in Fig. \ref{P_L2}.

Fig. \ref{P_L2} shows the effect of increasing the number of reflecting elements at the RIS when both the relay and the RIS are located at the mid point between the BS and the users circle center, in which case the channel in the second hop is weak. In this figure, we can clearly see that increasing the number of reflecting elements, the amount of reduced power is high in the presence of relay and low in its absence. Specifically, it can be seen that increasing the number of reflecting elements from 20 to 180 reduces the required power by 0.28 dBm if there is no relay (RIS-only), 3.94 dBm if there is a full-duplex relay with 5 antennas (the proposed model), and 4.9176 if there is a half-duplex relay. The presence of relay makes the BS-RIS-relay and relay-RIS-user much stronger than the BS-RIS and RIS-user channels,  which leads to increasing the effectiveness of each reflecting element.  In addition, the figure shows that the contribution of reflecting elements in half-duplex mode is higher than that in the full-duplex mode. This holds since the degree of freedom in the half-duplex mode is twice in that in the full-duplex mode. In other words, in half-duplex mode, we optimize each phase-shift twice, one for the first phase and the other for the second phase (i.e., optimize $\mathbf{\Theta_1}$ and $\mathbf{\Theta_2}$), while in the full-duplex mode, we are allowed to optimize each phase shift once. 

\begin{figure}[!t]
\centering
\includegraphics[width=3in]{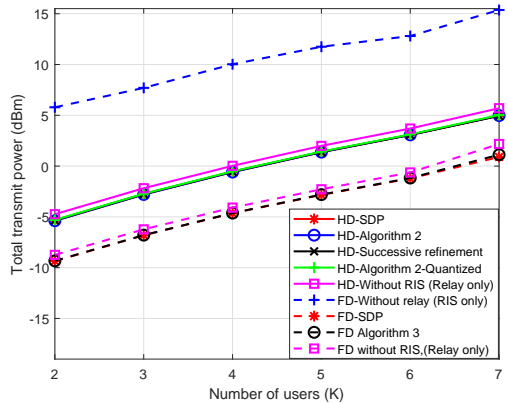}
\caption{The required total power versus the number of users in the system when $N=8$, $M=8$, $L=50$, and $R_{th}=2$.}
\label{P_K}
\end{figure}
Fig. \ref{P_K} shows that increasing the number of users in the system leads to increasing the required power to achieve an $R_{th}=2$. The figure confirms that the presence of a relay whether a half or a full-duplex one improves the system energy efficiency significantly. 

Fig. \ref{P_Itr} plots the convergence of the proposed algorithms in the half-duplex case. The figure shows that Algorithms \ref{algor1} and \ref{algor2} converge at around the sixth iteration. However, the convergence of the successive refinement approach depends on the number of reflecting elements. Although the figure shows that the SDP approach achieves the best result, each iteration requires solving two SDP problems, which leads to a high computational complexity compared to the other approaches.  


\section{Conclusion}\label{Sec:Conc}
This paper proposes the coexistence of relay and RIS to improve the energy efficiency of multiuser systems. The paper studies and optimizes the system with different cases such as half-duplex mode, full-duplex modes, continuous and discrete phase shifts. It proposes solutions for joint beamforming at the BS, the relay, and the RIS to minimize the required total power (at the BS and the relay) under given QoS at the users. Since the formulated optimization problems are difficult, we propose suboptimal solutions that rely on the fact that the equivalent channel from the BS to the users is weak relative to the relay-user channel. For optimizing the phase shifts, we propose two solutions for the continuous phase shifts and two other solution for the discrete phases. Some of the solutions are proposed to seek the best performance and others to seek the low complexity with negligible performance loss. In the simulation results, we evaluate and compare the contribution of both the relay and the RIS under changing different system parameters. In general, the presence of both RIS and relay improves the energy efficiency significantly. The results reveal that the full-duplex system outperforms all the other systems. They also show that the presence of a relay strengthen the contribution of the reflecting elements especially if the channels to the users are weak on average.

\bibliography{mylib}

\begin{thebibliography}{10}
\providecommand{\url}[1]{#1}
\csname url@samestyle\endcsname
\providecommand{\newblock}{\relax}
\providecommand{\bibinfo}[2]{#2}
\providecommand{\BIBentrySTDinterwordspacing}{\spaceskip=0pt\relax}
\providecommand{\BIBentryALTinterwordstretchfactor}{4}
\providecommand{\BIBentryALTinterwordspacing}{\spaceskip=\fontdimen2\font plus
\BIBentryALTinterwordstretchfactor\fontdimen3\font minus
  \fontdimen4\font\relax}
\providecommand{\BIBforeignlanguage}[2]{{%
\expandafter\ifx\csname l@#1\endcsname\relax
\typeout{** WARNING: IEEEtran.bst: No hyphenation pattern has been}%
\typeout{** loaded for the language `#1'. Using the pattern for}%
\typeout{** the default language instead.}%
\else
\language=\csname l@#1\endcsname
\fi
#2}}
\providecommand{\BIBdecl}{\relax}
\BIBdecl

\bibitem{obeed2021relay}
M.~Obeed and A.~Chaaban, ``Relay-reconfigurable intelligent surface cooperation
  for energy-efficient multiuser systems,'' \emph{arXiv preprint
  arXiv:2104.02849}, 2021.

\bibitem{alouiniR26}
E.~{Basar}, M.~{Di Renzo}, J.~{De Rosny}, M.~{Debbah}, M.~{Alouini}, and
  R.~{Zhang}, ``Wireless communications through reconfigurable intelligent
  surfaces,'' \emph{IEEE Access}, vol.~7, pp. 116\,753--116\,773, 2019.

\bibitem{cuiR22}
T.~J. Cui, M.~Q. Qi, X.~Wan, J.~Zhao, and Q.~Cheng, ``Coding metamaterials,
  digital metamaterials and programmable metamaterials,'' \emph{Light: Science
  \& Applications}, vol.~3, no.~10, pp. e218--e218, 2014.

\bibitem{WuR21}
Q.~{Wu} and R.~{Zhang}, ``Towards smart and reconfigurable environment:
  Intelligent reflecting surface aided wireless network,'' \emph{IEEE
  Communications Magazine}, vol.~58, no.~1, pp. 106--112, 2020.

\bibitem{9110889}
B.~{Di}, H.~{Zhang}, L.~{Song}, Y.~{Li}, Z.~{Han}, and H.~V. {Poor}, ``Hybrid
  beamforming for reconfigurable intelligent surface based multi-user
  communications: Achievable rates with limited discrete phase shifts,''
  \emph{IEEE Journal on Selected Areas in Communications}, vol.~38, no.~8, pp.
  1809--1822, 2020.

\bibitem{WuR1}
Q.~{Wu} and R.~{Zhang}, ``Intelligent reflecting surface enhanced wireless
  network via joint active and passive beamforming,'' \emph{IEEE Transactions
  on Wireless Communications}, vol.~18, no.~11, pp. 5394--5409, 2019.

\bibitem{huang2019reconfigurable}
C.~Huang, A.~Zappone, G.~C. Alexandropoulos, M.~Debbah, and C.~Yuen,
  ``Reconfigurable intelligent surfaces for energy efficiency in wireless
  communication,'' \emph{IEEE Transactions on Wireless Communications},
  vol.~18, no.~8, pp. 4157--4170, 2019.

\bibitem{9090356}
C.~{Pan}, H.~{Ren}, K.~{Wang}, W.~{Xu}, M.~{Elkashlan}, A.~{Nallanathan}, and
  L.~{Hanzo}, ``Multicell mimo communications relying on intelligent reflecting
  surfaces,'' \emph{IEEE Transactions on Wireless Communications}, vol.~19,
  no.~8, pp. 5218--5233, 2020.

\bibitem{8723525}
M.~{Cui}, G.~{Zhang}, and R.~{Zhang}, ``Secure wireless communication via
  intelligent reflecting surface,'' \emph{IEEE Wireless Communications
  Letters}, vol.~8, no.~5, pp. 1410--1414, 2019.

\bibitem{9087848}
Q.~U.~A. {Nadeem}, H.~{Alwazani}, A.~{Kammoun}, A.~{Chaaban}, M.~{Debbah}, and
  M.~S. {Alouini}, ``Intelligent reflecting surface-assisted multi-user miso
  communication: Channel estimation and beamforming design,'' \emph{IEEE Open
  Journal of the Communications Society}, vol.~1, pp. 661--680, 2020.

\bibitem{9226616}
P.~{Wang}, J.~{Fang}, X.~{Yuan}, Z.~{Chen}, and H.~{Li}, ``Intelligent
  reflecting surface-assisted millimeter wave communications: Joint active and
  passive precoding design,'' \emph{IEEE Transactions on Vehicular Technology},
  pp. 1--1, 2020.

\bibitem{WuR14}
Q.~{Wu} and R.~{Zhang}, ``Intelligent reflecting surface enhanced wireless
  network: Joint active and passive beamforming design,'' in \emph{2018 IEEE
  Global Communications Conference (GLOBECOM)}, 2018, pp. 1--6.

\bibitem{9194749}
Q.~U.~A. {Nadeem}, A.~{Chaaban}, and M.~{Debbah}, ``Opportunistic beamforming
  using an intelligent reflecting surface without instantaneous csi,''
  \emph{IEEE Wireless Communications Letters}, vol.~10, no.~1, pp. 146--150,
  2021.

\bibitem{9384319}
Q.~U.~A. {Nadeem}, A.~{Zappone}, and A.~{Chaaban}, ``Intelligent reflecting
  surface enabled random rotations scheme for the miso broadcast channel,''
  \emph{IEEE Transactions on Wireless Communications}, pp. 1--1, 2021.

\bibitem{HuangR3}
C.~{Huang}, A.~{Zappone}, G.~C. {Alexandropoulos}, M.~{Debbah}, and C.~{Yuen},
  ``Reconfigurable intelligent surfaces for energy efficiency in wireless
  communication,'' \emph{IEEE Transactions on Wireless Communications},
  vol.~18, no.~8, pp. 4157--4170, 2019.

\bibitem{sonR25}
E.~Bj{\"o}rnson and L.~Sanguinetti, ``Power scaling laws and near-field
  behaviors of massive mimo and intelligent reflecting surfaces,'' \emph{arXiv
  preprint arXiv:2002.04960}, 2020.

\bibitem{RIS_Relay}
M.~{Di Renzo}, K.~{Ntontin}, J.~{Song}, F.~H. {Danufane}, X.~{Qian},
  F.~{Lazarakis}, J.~{De Rosny}, D.~{Phan-Huy}, O.~{Simeone}, R.~{Zhang},
  M.~{Debbah}, G.~{Lerosey}, M.~{Fink}, S.~{Tretyakov}, and S.~{Shamai},
  ``Reconfigurable intelligent surfaces vs. relaying: Differences,
  similarities, and performance comparison,'' \emph{IEEE Open Journal of the
  Communications Society}, vol.~1, pp. 798--807, 2020.

\bibitem{ying2020relay}
X.~Ying, U.~Demirhan, and A.~Alkhateeb, ``Relay aided intelligent
  reconfigurable surfaces: Achieving the potential without so many antennas,''
  \emph{arXiv preprint arXiv:2006.06644}, 2020.

\bibitem{9225707}
Z.~{Abdullah}, G.~{Chen}, S.~{Lambotharan}, and J.~A. {Chambers},
  ``Optimization of intelligent reflecting surface assisted full-duplex relay
  networks,'' \emph{IEEE Wireless Communications Letters}, pp. 1--1, 2020.

\bibitem{7105651}
Z.~{Zhang}, X.~{Chai}, K.~{Long}, A.~V. {Vasilakos}, and L.~{Hanzo}, ``Full
  duplex techniques for 5g networks: self-interference cancellation, protocol
  design, and relay selection,'' \emph{IEEE Communications Magazine}, vol.~53,
  no.~5, pp. 128--137, 2015.

\bibitem{Boyd}
S.~Boyd, S.~P. Boyd, and L.~Vandenberghe, \emph{Convex optimization}.\hskip 1em
  plus 0.5em minus 0.4em\relax Cambridge university press, 2004.

\bibitem{cvx}
M.~Grant and S.~Boyd, ``Cvx: Matlab software for disciplined convex
  programming, version 2.1,'' 2014.

\bibitem{wu2018intelligent}
Q.~Wu and R.~Zhang, ``Intelligent reflecting surface enhanced wireless network:
  Joint active and passive beamforming design,'' in \emph{2018 IEEE Global
  Communications Conference (GLOBECOM)}.\hskip 1em plus 0.5em minus 0.4em\relax
  IEEE, 2018, pp. 1--6.

\bibitem{7547360}
Y.~{Sun}, P.~{Babu}, and D.~P. {Palomar}, ``Majorization-minimization
  algorithms in signal processing, communications, and machine learning,''
  \emph{IEEE Transactions on Signal Processing}, vol.~65, no.~3, pp. 794--816,
  2017.

\bibitem{8855810}
X.~{Yu}, D.~{Xu}, and R.~{Schober}, ``Miso wireless communication systems via
  intelligent reflecting surfaces : (invited paper),'' in \emph{2019 IEEE/CIC
  International Conference on Communications in China (ICCC)}, 2019, pp.
  735--740.

\bibitem{cui2014coding}
T.~J. Cui, M.~Q. Qi, X.~Wan, J.~Zhao, and Q.~Cheng, ``Coding metamaterials,
  digital metamaterials and programmable metamaterials,'' \emph{Light: Science
  \& Applications}, vol.~3, no.~10, pp. e218--e218, 2014.

\bibitem{kaina2014shaping}
N.~Kaina, M.~Dupr{\'e}, G.~Lerosey, and M.~Fink, ``Shaping complex microwave
  fields in reverberating media with binary tunable metasurfaces,''
  \emph{Scientific reports}, vol.~4, no.~1, pp. 1--8, 2014.

\bibitem{NadeemR4}
Q.~{Nadeem}, A.~{Kammoun}, A.~{Chaaban}, M.~{Debbah}, and M.~{Alouini},
  ``Asymptotic max-min sinr analysis of reconfigurable intelligent surface
  assisted miso systems,'' \emph{IEEE Transactions on Wireless Communications},
  pp. 1--1, 2020.

\end{thebibliography}

\bibliographystyle{IEEEtran}

%
%
%
%
%
%
%
%

\end{document}